\documentclass[a4paper]{IEEEtran}

\IEEEoverridecommandlockouts
\usepackage{multicol}
\usepackage{graphicx}
\usepackage{amssymb, amsmath, amsfonts, amsthm}
\usepackage{multirow}
\usepackage{bm}
\usepackage{nomencl}
\usepackage{mcite}
\usepackage[noadjust]{cite}
\usepackage{setspace}
\usepackage{flushend}
\usepackage{xcolor}
\usepackage{acronym}
\usepackage{nomencl}
\usepackage{etoolbox}
\usepackage{dblfloatfix}
\usepackage{placeins}
\acrodef{ibr}[IBR]{Inverter-Based Resource}
\acrodef{gfor}[GFM]{Grid-Forming}
\acrodef{gfol}[GFL]{Grid-Following}
\acrodef{sg}[SG]{Synchronous Generator}
\acrodef{vsb}[VSB]{Voltage Source Behaviour}
\acrodef{dsi}[DSI]{Dynamic Similarity Index}
\acrodef{vsbi}[VSbI]{Voltage Source behind an Impedance}
\acrodef{tso}[TSO]{Transmission System Operator}
\acrodef{pll}[PLL]{Phased-Locked Loop}
\acrodef{mimo}[MIMO]{Multiple-Input Multiple-Output}
\acrodef{vsm}[VSM]{Virtual Synchronous Machine}
\acrodef{poc}[POC]{Point of Connection}
\acrodef{tf}[TF]{Transfer Function}
\acrodef{scr}[SCR]{Short-Circuit Ratio}

\begin{document}

\title{A Dynamic Similarity Index for Assessing Voltage Source Behaviour in Power Systems}
%Option 1: Voltage Behavior Analysis in Large Power Electronic Dominated Systems: A Comprehensive Index Approach
%Option 2: Voltage Unification: An Index-Based Analysis of Dynamic Similarities in Converters
%Option 3: Dynamic Equivalence: Assessing Voltage Behavior Uniformity in Power System Converters
%Option 4: Dynamic Equivalence: Assessing Voltage Behavior\\ Similarity in Power System Converters
%Option 5: Quantifying Dynamic Response Similarity in Inverter-Based Generation for System Studies by Using Dynamic Similarity Index
% Dynamic Similarity Index: A Tool for Assessing Voltage Dynamics in Modern Power Systems
\author{Onur~Alican,~\IEEEmembership{Student Member,~IEEE,}
Dionysios~Moutevelis,~\IEEEmembership{Member,~IEEE},
Josep~Ar\'{e}valo-Soler,~\IEEEmembership{Student Member,~IEEE}, Carlos Collados-Rodriguez,~\IEEEmembership{Member,~IEEE}, Jaume Amor\'{o}s, Oriol~Gomis-Bellmunt,~\IEEEmembership{Fellow,~IEEE}, \\
Marc~Cheah-Mañe,~\IEEEmembership{Member,~IEEE}, 
Eduardo~Prieto-Araujo,~\IEEEmembership{Senior Member,~IEEE}}
        
        % <-this % stops a space
% \thanks{M. Shell was with the Department
% of Electrical and Computer Engineering, Georgia Institute of Technology, Atlanta,
% GA, 30332 USA e-mail: (see http://www.michaelshell.org/contact.html).}% <-this % stops a space
% \thanks{J. Doe and J. Doe are with Anonymous University.}% <-this % stops a space
% \thanks{Manuscript received April 19, 2005; revised August 26, 2015.}

\maketitle

\begin{abstract}

\textcolor{black}{Due to the fundamental transition to a power electronic dominated power system, the increasing diversity of dynamic elements underscores the need to assess their similarity to mature electrical engineering models.}
This article addresses the concept of the Dynamic Similarity Index (DSI) for its use in, power electronics-dominated networks.
The DSI is a multi-purpose tool developed to be used by different stakeholders (e.g., converter manufacturers and system operators). 
%
% The concept of the DSI is funded on computing the error between two dynamical systems in the frequency domain using the Singular Value Decomposition (SVD) for Multiple Input Multiple Output (MIMO) systems.
%
 Such an index is calculated per frequency, which serves to anticipate \textcolor{black}{potential differences} in particular frequency ranges of interest between the model under study and the reference model.
Within the scope of this study, the dynamic
similarity of inverter-based generators to an ideal voltage source
behind an impedance is \textcolor{black}{assessed}, due to the relevance of this
fundamental circuit in the representation of generation units in
power system studies.
The article presents two potential applications based
on this mathematical framework.
\textcolor{black}{First, for manufacturers to evaluate control performance compared to a reference model and second, it enables operators to diagnose buses with voltage vulnerability based on a user-defined reference \ac{scr} value. The DSI results for these 
two case studies are validated using Matlab Simulink simulations.} 

\end{abstract}


\IEEEpeerreviewmaketitle

\section{Introduction}
\label{sec.intro}
%\section{Voltage source behaviour}
\label{sec:intro}
\subsection{Motivation}
\label{sec:motivation}

Traditional power systems exhibit high \ac{vsb} due to the \ac{sg} capability to impose system voltage and phase angle~\cite{kundur}.
Also, since the impedances of \ac{sg} stators are not negligible, \acp{sg} are typically represented as voltage sources behind an impedance in power system stability and converter integration studies~\cite{sallam_2015_power}.
However, due to the increasing need for decarbonization,
%in modern power systems,
\ac{sg}s are gradually
%being phased out and
replaced by \acp{ibr}.
%Such a shift alters the dynamics of the power system.
Conventional controllers of such IBRs often exhibit limited \ac{vsb}, compared to \acp{sg}, which leads to increased grid voltage variations against disturbances~\cite{2017_high,2021_global}.
%
%This shift reduces the  of the system hence, voltage becomes more vulnerable to disturbances 
% %
%     Voltage source behavior is crucial in modern power systems with high penetration of renewable energy sources. 
%  Power grids with a high penetration of renewables may experience voltage stability issues and reduced resilience to disturbances (Wang et al., 2023; Al-Zahrani et al., 2020).
% %
The ideal \ac{vsb} of a device consists in maintaining constant both the magnitude and the phase angle of the voltage at its connection point under any disturbance.
This requires the capability to instantaneously inject or absorb any amount of current during the disturbance, something unfeasible in practice~\cite{moutevelis2023taxonomy}.
For this reason in practical scenarios, \acp{tso} may require a \ac{vsbi} characteristic in a frequency range of 5 Hz to 1 kHz to avoid detrimental controller interactions as well as to maintain voltage stability~\cite{PAOLONE,2019_high}. 
In this context, the capability to quantify the dynamic similarity of various generators, both conventional and inverter-based, to an ideal \ac{vsbi} characteristics is highly desirable for both component control design and system-wide analysis.
%Having a system that is not redundant is not desired by the \ac{tso} hence they require certain characteristics for generation units such as \ac{vsbi} behavior. The reason behind such a desire is that an ideal voltage source can inject any amount of current to keep the voltage constant at load disturbances, hence seeking such an ideal unit is not realistic. So, it is crucial to have a \ac{vsbi} characteristic in a frequency range of 5 Hz to 1 kHz to avoid detrimental controller interactions as well as maintain voltage stability~\cite{PAOLONE,2019_high}. 
\subsection{Literature Review}
\label{sec:literature}
The resemblance of \acp{ibr} to a \ac{vsbi} with regards to their dynamic operation highly depends on their control structure.
The \ac{ibr} controllers are typically categorized between \ac{gfol} and \ac{gfor}, based on their capability, or lack thereof, to regulate the voltage and frequency at their connection point.
%
%In general, introducing more \ac{gfol}-\acp{ibr}, lowers the \ac{vsb} of the system for the following reasons.
%
%Introducing more \acp{ibr} that operate in \ac{gfol} mode, lowers the \ac{vsb} of the system.
%
\ac{gfol} converters use a \ac{pll} that tracks the voltage
of the grid and estimates its angle, which is then used in the converter control.
%
% Unlike \acp{sg}, \ac{gfol} converters do not provide immediate reaction to load or generation variations due to 
% the (intrinsically) delayed response of the \ac{pll}~\cite{pll_slow}.
%
\textcolor{black}{This leads to potential instabilities of \ac{gfol} converters under weak grid conditions~\cite{pll_weak}. 
Moreover, since \ac{pll} tracks the grid voltage and estimate angle, \ac{gfol} converters are intended to inject current into the grid within its limits~\cite{wang2020grid}.}
% Additionally, \ac{gfol} converters typically lack an explicit voltage controller and for most applications, directly controlling the current injection at their connection point is preferred.} 
%
For the above reasons, \ac{gfol} converters are usually represented by current sources in sub-transient time frames~\cite{rosso2021grid,esig}.
%since they control the current ~\cite{rosso2021grid,esig}.
%
\textcolor{black}{However, this representation is not unique, since \ac{gfol} converters can be assembled with various control strategies that have different tuning options. 
Therefore, quantifying the dynamic similarity between \ac{vsbi} and \ac{gfol}-\acp{ibr} is of interest.}

% and the quantification of the dynamic similarity between a \ac{vsbi} and \ac{gfol}-\acp{ibr} is of interest since they can be equipped with various control blocks.
%
%Accordingly, introducing more \ac{gfol}-\acp{ibr} might make the system voltage more susceptible to variations and oscillations after a disturbance, hence lowering the \ac{vsb} of the overall system~\cite{green_stability}.
%

%Contrary to \ac{gfol} converters, 
\ac{gfor} converters 
%do not need a \ac{pll} to remain synchronized with the grid.
%however PLL may be used for reference measurement.
%Such converters 
are capable of imposing the voltage and angle at their connection bus by means
of a dedicated voltage controller.
%allowing their connection to weak grids.
%and for providing various grid services. %in a way similar to the \acp{sg}.
%
For this reason, they are usually represented as \ac{vsbi} from the grid perspective in the sub-transient to the transient time frame~\cite{esig}.
%
%That capability implies voltage source behavior and provides a fast response to load and generation variations in the system~\cite{rosso2021grid}.
%
%Several types of \ac{gfor} control have been proposed in the literature such as droop control, \ac{vsm} and Virtual Oscillator Control (VOC)~\cite{gfor}. 
%
%These characteristics of the \ac{gfor} make them suitable for connection to weak grids and for providing various grid services.
% , such as black-start, frequency and voltage regulation, and power oscillation damping~\cite{rodriguezamenedo,harrison2023demystifying}.
%
Due to the different available control topologies (e.g., droop control, \ac{vsm} and Virtual Oscillator Control (VOC)~\cite{gfor}) and the wide range for the controller parameter selection, \textcolor{black}{quantifying the level of \ac{vsbi} behaviour} of each \ac{gfor}-\ac{ibr} exhibits is not straightforward.
%    However, putting too many \ac{gfor} converters to ensure voltage stability may cause interactions between those converters which can lead to instability.
%

The \ac{vsb} of generation units is also indicative of potential interactions between them.
Generation units with high \ac{vsb}, which are located electrically close or connected to a strong grid, may induce oscillations 
and interact between them or with the network, leading to instability \cite{Zhao2022,Zhao2023}.
For this reason, the ability to quantify the \ac{vsbi} of each generation unit, depending on their control mode and control parameter selection, would be beneficial for its control design and for system-level planning.
%
%In other words, units that are operating in different modes with various tuning will not have the same dynamic voltage behavior. So, it is crucial to be able to assess the \ac{vsb} of the units across a range of frequencies.
%

%There are few works on assessing the voltage source behavior of generation units. 
%
In the literature, there have been limited studies that systematically quantify the \ac{vsb} of different generation units.
In \cite{Xu2023}, the authors propose a voltage stiffness index that considers the impedance and phase angle between the network and the different interconnected devices at the fundamental frequency.
However, it would be beneficial to asses the \ac{vsb} of generators across a wide frequency range.
In references~\cite{Shah2022,freq_gfol_gfor}, the authors describe a frequency-scan-based characterization of \ac{gfor} and \ac{gfol} modes.
However, an analytical link between the oscillation modes and the \ac{vsb} of the generation units is not included.
%
%However, the analysis is shallow and lacks a mathematical link for voltage source behavior.
%
Later in~\cite{VATTAKKUNI20236042}, the authors compare the \ac{vsb} of various \ac{gfor} inner control techniques with a \ac{vsbi} in both time and frequency domains.
In this study, \ac{gfol} converters are not considered.
Finally, in \cite{vsb_linbin} the voltage source behavior of \ac{gfor} and \ac{gfol} are mathematically compared by using singular values.
Also, the effect of different control topologies on the voltage source behavior of \ac{gfor} is shown. 
However, various tuning options of \ac{gfol} and its effect on the \ac{vsb} is not studied.

The similarity between different dynamical systems can be observed in several ways. 
%
%The Buckingham $\pi$-Theorem simplifies systems through dimensionless parameters, but it becomes limited for large or complex systems with many variables~\cite{buckingham_book}.
%
Modal analysis fully captures the damping ratio and frequency of all the system modes~\cite{modal}.
However, when the order of the system is large, inspecting all the system modes is not straightforward.
%
%each mode's damping ratio and frequency however, it is compelling when having large or different numbers of modes between systems. 
%
Applying a perturbation and comparing the time-domain signals of different systems is an alternative way to assess the dynamic similarity between them.
The shortcoming of this approach is that the different dynamics of the system are unequally excited from each disturbance, limiting the accuracy of the observed results for the full frequency spectrum of interest.
%might not contain all frequencies of interest hence introducing inaccuracies.
%
Finally, assessing the dynamic similarity in the frequency domain provides information across the full dynamical spectrum.
However, the typical representation of power system devices using rotating reference frames in the frequency domain consists of defining a 2$\times$2 \ac{mimo} system, leading to four frequency domain plots to observe the full dynamic operation of the studied systems~\cite{machowski_2011_power}.
%in electrical engineering, the systems are mostly studied in $qd$ reference frame which makes 4 frequency domain plots to observe .
%
The proposed \ac{dsi} method overcomes all these issues by
%taking the error between two dynamical systems in the frequency domain and, by using maximum singular values, the
quantifying the maximum amplification of the error
%can be observed
in a single plot at all frequencies, within a rotating reference frame modelling approach.
%

% % Additionally, a systematic algorithm for quantifying the \ac{vsb} is missing.

%  % A voltage similarity calculation for voltage loss analysis is performed in \cite{voltage_sim} and \cite{2017_topology} by using correlation analysis and discrete Fréchet distance, respectively. However, none of the studies reflect 

% %So, it is clear that there is no index that provides insight into how much different generation units exhibit VSB  across a frequency range. Therefore,
% \subsection{Contribution}
% \label{sec:contribution}
% %
% In this paper, a \ac{dsi} is proposed for the calculation of the dynamic similarity between two systems in the small-signal region and across a frequency range.
%This index is 

\subsection{Contribution}
In this paper, \textcolor{black}{the \ac{dsi} concept is introduced for different applications.} By selecting \ac{vsbi} as a reference, \textcolor{black}{note any reference can be selected depending on the study) first, the \ac{vsb} of both \ac{gfol} and \ac{gfor} converter with different control configurations is shown.} Then, the index is applied on the system level to identify which parts of the network \textcolor{black}{has different value with respect to the reference}.
The contributions of this paper can be summarized as follows:
\begin{itemize}
    %
    % \item The dynamic behavior of a voltage source behind an impedance is analytically derived using impedance analysis in the Laplace domain.
    %
    \item A novel metric, (\ac{dsi}), \textcolor{black}{is presented to evaluate} the dynamic similarity between two systems, is proposed. The \ac{dsi} is calculated in the \textcolor{black}{frequency domain} in MIMO systems.
    \item The proposed index is applied to show the effect \textcolor{black}{that} various operational modes and control parameter selections of \acp{ibr}
    \textcolor{black}{have on} their \ac{vsb} assessment. Such an application can be useful for converter manufacturers or consultancy companies to reveal the impact of the converter control setup on the network operation.
    \item \textcolor{black}{Finally, the \ac{dsi} is used for the identification of buses that have high/low VSB in a 100\% \ac{ibr} based network.  This information can help TSOs determine critical connection points for newly-commissioned \acp{ibr} and their suitable control structures.}
    \end{itemize}
\textcolor{black}{Note that, such applications are just example usages of the methodology. Other operations of the methodology might arise.}
The results are supported by simulations for two case studies based on well-known benchmarks, \textcolor{black}{including the IEEE 9-bus to illustrate the concept and 118-bus systems showcase it in a larger network.} 
\subsection{Paper Organization}
\label{sec:organization}
The organization of the rest of the paper is the following.
The \ac{vsbi} behavior is derived analytically and studied in the frequency domain in Section~\ref{sec.vsb}.
In Section~\ref{sec.dsi}, the \ac{dsi} and its applications are introduced.
Section~\ref{sec.case_study} presents the case studies while Section~\ref{sec.conc} concludes the paper and suggests future research directions.
\section{DSI Reference Model Selection} \label{sec.vsb}
%for Small Perturbations}
%
\textcolor{black}{This section aims to provide insights for the reference selection, since this is critical for assessing the dynamic similarity of a system.}
\textcolor{black}{Note that, in this work, the reference selection is based on a model of a \ac{vsbi}.}
To this end, the small-signal dynamic characteristics of a \ac{vsbi} are first analytically derived by using impedance analysis.
%
%Moreover, the $X/R$ to damping ratio ($\xi$) derivation is studied in the Appendix (Section \ref{sec.appen1}) by using $R$ and $L$. 
%
Then, an example system, consisting of two voltage sources, \textcolor{black}{one representing the system under study and the other external grid},
%
%one system under study and the other representing the rest of the network is represented.
%
%Such a system is shown 
is used to illustrate the effect of the \ac{vsbi} parameters on the overall system dynamic performance.
%
%effect of the system under study on the voltage dynamics while maintaining the rest of the network.
%a system under study and the other representing the rest of the network.
%
%Later the impact of the series impedance of the system under study on the dynamic characteristics of the voltage at the connection point is highlighted in the frequency domain by changing the series impedance.
%
The specification and analytic derivation of the desired dynamic characteristics of a voltage source will then be leveraged in Sections~\ref{sec.dsi} and~\ref{sec.case_study} to be used as a reference for the DSI calculation.
%and to compare it against the dynamic performance of converter-based generators.
%
\subsection{Impedance Analysis Preliminaries}
\label{sec.impedance}
%In this section, the small-signal framework for stability analysis is briefly presented. In Section~[0], it will be used for the calculation of the \ac{dsi}.
%
The starting point is the representation of a power system as a set of nonlinear ordinary differential equations in the form of~\cite{kundur}
\begin{equation}  
\label{eq:ode}
\begin{aligned}
\frac{d \boldsymbol{x}}{dt}
=
\boldsymbol{f(x,u)}
\end{aligned}
\end{equation}
%
%where $x(t)$, $y(t)$, $u(t)$ are the state, output and input vectors of the system, respectively.
%
where $ \boldsymbol{x}= \boldsymbol{x(t)}: \mathbb{R}^{n}$
and $\boldsymbol{u}=\boldsymbol{u(t)}: \mathbb{R}^{m}$ are column vectors of state and input variables, respectively,
$\boldsymbol{ f} : \mathbb{R}^{n+m} \rightarrow \mathbb{R}^{n}$ is a nonlinear function and $n$, $m$  are the number of states and inputs, respectively.
One should note that the representation of~\eqref{eq:ode} is general and can always be achieved
%when the line dynamics are not neglected
~\cite{canizares1998calculating,chen2021improving}.
By linearizing \eqref{eq:ode} and selecting $k$ number of outputs,
%applying the Taylor series approximation without considering the second order term of $\boldsymbol{f}$ and selecting $k$ number of outputs,
the linear state-space model of the system is represented as follows~\cite{ogata2010modern}
\begin{equation}
\label{eq:linear_ode}
\begin{aligned}
\frac{d\boldsymbol{\Delta x}}{dt}=\boldsymbol{A} \boldsymbol{\Delta x}+ \boldsymbol{B} \boldsymbol{\Delta u}
\\    \boldsymbol{\Delta y}=\boldsymbol{C} \boldsymbol{\Delta x} + \boldsymbol{D} \boldsymbol{\Delta u}
\end{aligned}     
\end{equation}
where $ \boldsymbol{y}= \boldsymbol{y(t)}: \mathbb{R}^{k}$ is the output column vector, $\boldsymbol{A} \in \mathbb{R}^{n \times n}$, $\boldsymbol{B} \in \mathbb{R}^{n \times m}$, $\boldsymbol{C} \in \mathbb{R}^{k \times n}$, and $\boldsymbol{D} \in \mathbb{R}^{k \times m}$ are the state, input, output and feedforward matrices, respectively. From the state-space representation, the \acp{tf} between the system inputs and outputs in the Laplace domain are derived using
  \begin{equation}
\label{eq.ABCD2imp}
     \frac{\boldsymbol{y}(s)}{\boldsymbol{u}(s)}=\boldsymbol{C}(s\boldsymbol{I}-\boldsymbol{A})^{-1}\boldsymbol{B}+\boldsymbol{D}
\end{equation}
where $s$ is the Laplace operator, and $\boldsymbol{u}(s)$, $\boldsymbol{y}(s)$ are the input and output functions, respectively, in the Laplace domain. %Such selections are crucial for studying the desired dynamics.
When these inputs/outputs refer to the incremental voltages and currents~(i.e., $\boldsymbol{y}(s)=\Delta \boldsymbol{v}$, $\boldsymbol{u}(s)=\Delta \boldsymbol{i}$), the expression in~\eqref{eq.ABCD2imp} defines the system impedance matrix $\boldsymbol{Z(s)}$~\cite{sun2011impedance}.
%
 % For the cases where the inputs and outputs of the system are electrical variables (voltages and currents), the \ac{tf} matrix $\boldsymbol{Z(s)}$ of the system is also called impedance matrix~\cite{sun2011impedance}.
 %
 This matrix contains all the dynamics of the linearized system from \eqref{eq:linear_ode}, including those of the control and the electrical components~(e.g., passive filters, transmission lines, etc.).
 The dual expression of the impedance matrix is called the admittance matrix and can be both expressed as
%This matrix describes the relation between the incremental inputs/outputs~(voltages/currents)  $\Delta \boldsymbol{v} = [v_q \; v_d]^\intercal$ and $\Delta \boldsymbol{i} = [i_q \; i_d]^\intercal$~($^{\intercal}$ being the transpose operator) of a system in a rotating $qd$ reference frame, as follows:
%
\begin{equation}
\label{eq.imp}
    \begin{aligned}        
    \Delta \boldsymbol{u}=\boldsymbol{Z(s)}\Delta \boldsymbol{i} \\
   \Delta \boldsymbol{i}=\boldsymbol{Y(s)}\Delta \boldsymbol{u}
    \end{aligned} 
\end{equation}
where $\boldsymbol{Y(s)}=\boldsymbol{Z(s)}^{-1}$ is the admittance matrix and $\Delta \boldsymbol{u} = [u_q \; u_d]^\intercal$, $\Delta \boldsymbol{i} = [i_q \; i_d]^\intercal$~($^{\intercal}$ being the transpose operator) are the incremental inputs/outputs~(voltages/currents) of a system in a rotating $qd$ reference frame.
From~\eqref{eq.imp}, it can be observed that when the elements of $\boldsymbol{Z(s)}$ have small values~(elements of $\boldsymbol{Y(s)}$ have high values), a change in the current causes a small change in the voltage.

This dynamic behavior resembles the behavior of an ideal voltage source, for which the voltage remains constant independently of the current it provides.
Hence, when \eqref{eq.imp} models generators, it can be considered that a generator with low element values of \boldsymbol{$Z(s)$}~(high element values of \boldsymbol{$Y(s)$}) exhibits higher voltage behavior than a source with high element values of \boldsymbol{$Z(s)$}~(low element values of \boldsymbol{$Y(s)$}).
In the remaining of this section, impedance analysis will be used to analytically derive the dynamic characteristics of voltage sources behind an impedance.
%
%In Section~\ref{sec.dsi}, the impedance matrix will be used for the calculation of the \ac{dsi} and the \ac{vsb} assessment of various generators.
%
%will be shown as small changes in the voltage hence more voltage source. So we can say,  the smaller \boldsymbol{$Z(s)$} and the higher $Y(s)$ the more voltage source behavior. It can be seen that with $Z_f$ 0.07  the magnitude of the admittance matrix is higher than $Z_f$ 0.7 so it has more voltage source behavior.
%
%
%
%
\begin{figure}[!t]
\centering
\includegraphics[width=0.8\columnwidth]{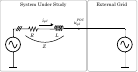}
\vspace{-0.3cm}
\caption{\textcolor{black}{Single-line diagram of a simple radial system with 2 voltage sources
behind an impedance. Z (System Under Study) and (External Grid)}}
\label{fig_th_case}
\end{figure}
\begin{figure}[!t]
\centering
\includegraphics[width=1\columnwidth]{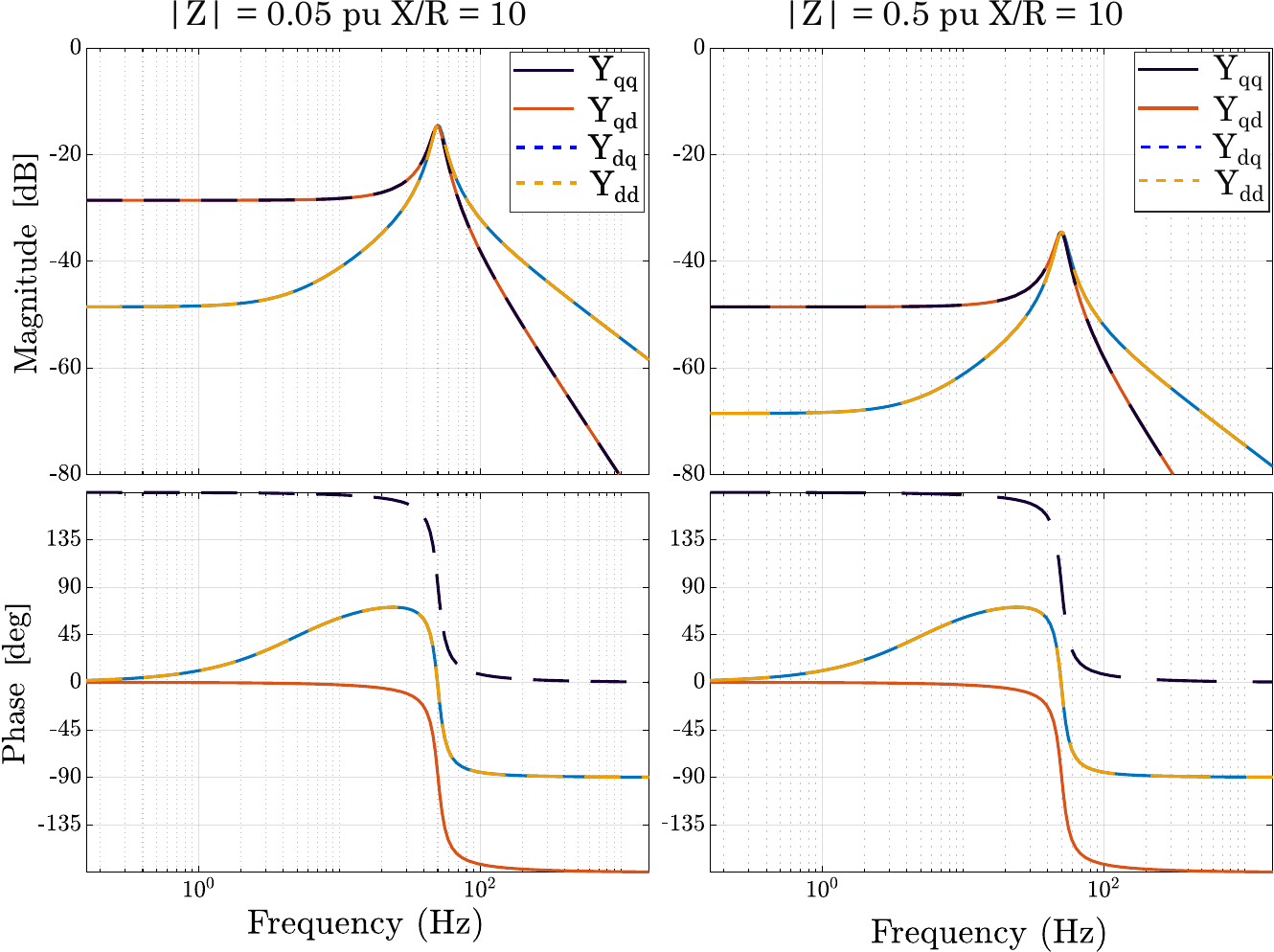}
\caption{Bode diagrams of the admittance matrix of the system under study with $|Z|$= 0.05 pu (left) and $|Z|$= 0.5 pu (right) with $X/R$ ratio of 10.}
\label{fig_bode_sc1}
\end{figure}
\subsection{Essential Analysis of \ac{vsbi} Behaviour}
%\subsection{Derivation of Voltage Source Behaviour}
\label{sec.derivation}
The differential equations that describe the current $i_{q,d}$ flowing through a resistor $R$ connected in series with an inductor $L$ in a $qd$ reference frame can be written as
\begin{equation}  
\begin{aligned}
\label{eq:current}
    u_q^{POC}(t)=i_q(t)R+L\frac{di_q(t)}{dt}+\omega_0 Li_d(t)
\\
    u_d^{POC}(t)=i_d(t)R+L\frac{di_d(t)}{dt}-\omega_0 Li_q(t)
\end{aligned}
\end{equation}  
where $u_{q,d}^{POC}$ are the voltage components at the \ac{poc} and $\omega_0=2 \pi f_0$ is the nominal frequency of the network.
%and $R$ and $L$ are the resistance and inductance components of $Z_1$ respectively.
%
By using~\eqref{eq:linear_ode}, \textcolor{black}{where time dependency is removed for simplicity},  \eqref{eq:current} is rewritten as follows
\begin{equation}
\begin{aligned}
        \frac{d}{dt} 
        \begin{bmatrix}
        \Delta i_q\\
        \Delta i_d
    \end{bmatrix} 
    =
    \underbrace{
    \begin{bmatrix}
        -\frac{R}{L} & -\omega_0\\
        \omega_0 & -\frac{R}{L}\\
    \end{bmatrix}}_\text{$\boldsymbol{A}$}
    \underbrace{
     \begin{bmatrix}
        \Delta i_q\\
        \Delta i_d
    \end{bmatrix}}_\text{x}
    +
    \underbrace{
     \begin{bmatrix}
        -\frac{1}{L} &0\\
        0 & -\frac{1}{L}
    \end{bmatrix}}_\text{$\boldsymbol{B}$} 
    \underbrace{
     \begin{bmatrix}
        \Delta u_q^{POC}\\
        \Delta u_d^{POC}\\
    \end{bmatrix}}_\text{$\boldsymbol{u}$} \\
% \end{equation}
% \begin{equation}
% \begin{split}
\underbrace{
    \begin{bmatrix}
       \Delta i_q\\
       \Delta i_d
    \end{bmatrix}}_\text{$\boldsymbol{y}$}
    =
    \underbrace{
    \begin{bmatrix}
        1 & 0\\
        0 & 1\\
    \end{bmatrix}}_\text{$\boldsymbol{C}$}
    \begin{bmatrix}
       \Delta i_q\\
       \Delta i_d
    \end{bmatrix}
    +
    \underbrace{
     \begin{bmatrix}
        0 & 0 \\
        0 & 0 
    \end{bmatrix}}_\text{$\boldsymbol{D}$} 
    \begin{bmatrix}
       \Delta u_q^{POC}\\
        \Delta u_d^{POC}\\
    \end{bmatrix}  
% \end{split}
\label{eq_mat_ABCD}
\end{aligned}
\end{equation}
By substituting matrices $\boldsymbol{A}$, $\boldsymbol{B}$, $\boldsymbol{C}$, and $\boldsymbol{D}$ from \eqref{eq_mat_ABCD} into \eqref{eq.ABCD2imp}, the admittance expression is written as
% %
% \begin{align} \label{sI-A}
%     \begin{bmatrix}
%         s & 0\\
%         0 & s
%     \end{bmatrix}
%     -
%     \begin{bmatrix}
%         -\frac{R}{L} & -\omega_0\\
%         \omega_0 & -\frac{R}{L}\\
%     \end{bmatrix}   
%     =
%      \begin{bmatrix}
%           s+\frac{R}{L} & \omega_0\\
%         -\omega_0 & s+\frac{R}{L}\\
%     \end{bmatrix}     
% \end{align}
% %
% Taking the inverse of \eqref{sI-A}, the following expression is obtained:
% %
% \begin{align}
%     \frac{1}{(s+\frac{R}{L})^2-(-\omega_0)^2}
%      \begin{bmatrix}
%           s+\frac{R}{L} & \omega_0 \\
%         -\omega_0 & s+\frac{R}{L}\\
%     \end{bmatrix}.      
% \end{align}
% Finally, multiplying \eqref{sI-A} by $\boldsymbol{C}$ and $\boldsymbol{B}$ and adding $\boldsymbol{D}$, the $2 \times 2$  admittance matrix can be written as:
%
\begin{align}
\label{eq:vsbi}
     \begin{bmatrix}
        \Delta i_q\\
        \Delta i_d
    \end{bmatrix}
    =
    \frac{1}{s^2+\frac{2sR}{L}+\frac{R^2+\omega_0^2L^2}{L^2}}
        \begin{bmatrix}
          s+\frac{R}{L} & \omega_0 \\
          -\omega_0 & s+\frac{R}{L}
    \end{bmatrix}
         \begin{bmatrix}
        \Delta u_q^{POC}\\
        \Delta u_d^{POC}\\
    \end{bmatrix}.
\end{align}
It can be seen from \eqref{eq:vsbi} that second-order \acp{tf} \textcolor{black}{obtained}, despite the admittance having a first-order \ac{tf} in each separate axis of the $qd$ formulation.
%
%This is important because \dots
%
Equation~\eqref{eq:vsbi}, \textcolor{black}{which is expressed in $RL$ parameters,} defines the dynamics of an ideal voltage source behind an impedance in the Laplace domain.
% (0).
%
A mapping from $RL$ parameters to the equivalent damping and reactance-to-resistance ratios ($\xi$ and $X/R$, respectively) is provided in the Appendix.
\textcolor{black}{Such analytical mapping provides insight into selecting the damping ratio of the user-defined reference.}
%
%Moreover, the $X/R$ to damping ratio ($\xi$) derivation is studied in the Appendix (Section \ref{sec.appen1}) by using $R$ and $L$.
%
In the following Sections, a metric \textcolor{black}{called \ac{dsi}} is proposed to quantify how closely potential converter control structures (both \ac{gfol} and \ac{gfor}) resemble the dynamic behavior defined in~\eqref{eq:vsbi} across different frequency ranges.
%
% \subsection{Illustrative Example}
% \label{sec.illustrative}

To further illustrate the effect of the impedance parameter values on the dynamic behaviour of a \ac{vsbi}, an example system is used, shown in Fig.~\ref{fig_th_case} consisting of two Th\'{e}venin circuits.
For the purpose of this analysis, two cases are considered.
First, a small value for the impedance \textcolor{black}{$|Z|$} is chosen equal to 0.05 pu while for the second case, its value is increased to 0.5 pu.
%, and note that only system 1 is studied and shown in the frequency domain.
%
For both cases, an $X/R$ ratio of 10 is chosen for \textcolor{black}{$Z$}.
Fig. \ref{fig_bode_sc1} shows the Bode diagram of the admittance matrix of impedance \textcolor{black}{$Z$}. When \textcolor{black}{$|Z|$} is set to 0.05 pu, a higher peak resonance can be observed compared to the value of 0.5 pu. 
This can be interpreted as presenting higher voltage source behavior.
%
% The rationale behind it is the following.
%
Also, since $\omega_0$ is chosen to be 314.15 rad/s, \textcolor{black}{the current will have a peak resonance in 50 Hz due to the peak observed in the Bode diagram.} 
\textcolor{black}{From the above analysis, it can be concluded that the selection of the impedance affects the \ac{vsb} of the reference.  By using the derived analytical expressions, and damping to $X/R$ ratio (see Appendix) the user can customize its reference selection which will be used in \ac{dsi} calculation.}
\section{Dynamic Similarity Index}
\label{sec.dsi}
\subsection{Proposed Method}
\label{sec.dsi_overview}
\begin{figure}[!t]
\centering
\includegraphics[width=1\columnwidth]{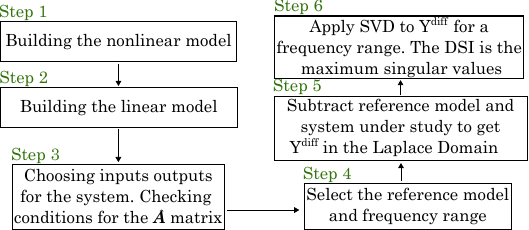}
\vspace{-0.3cm}
\caption{Flowchart of the methodology}
\label{fig_flow}
\end{figure}

In this section, \textcolor{black}{the \ac{dsi}} is proposed to quantify the resemblance of a given dynamic system to a \ac{vsbi}.
%, which is the main contribution of this paper.
%
An overview of the proposed approach can be seen in Fig.~\ref{fig_flow}.
First, the nonlinear model of the system is defined, as in \eqref{eq:ode}.
The analytical modelling of the different power system elements in the form of~\eqref{eq:ode} can be found in the literature~\cite{moutevelis2022bifurcation,collados}.
Then, the nonlinear model is linearized around one operation point, resulting in the linear model of the system, as in \eqref{eq:linear_ode}.
%~\cite{markovic2021understanding}.
%
%The linear model of the full system is validated against the nonlinear one by applying a small disturbance to ensure that the small-signal dynamics are properly captured.
%
\textcolor{black}{Note that, to ensure the integrity of the work, the methodology for calculating the \ac{dsi} is explained starting from the development of both nonlinear and linear models. However, the methodology itself is flexible and can be applied to any linear model, regardless of how it was derived.}

\begin{figure}[!t]
\centering
\includegraphics[width=1\columnwidth]{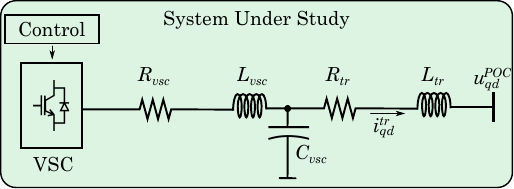}
\vspace{-0.3cm}
\caption{\textcolor{black}{System under study} by choosing $u_{qd}^{POC}$ (as input) and $i_{qd}^{tr}$ (as output) as an example}
\label{fig_gu_sys_slit}
\end{figure}

\textcolor{black}{After obtaining the linear model}, the system is separated based on the input and output selection~\cite{sun2011impedance}.
This results in two subsystems, one representing the dynamics of the system under study~(\textcolor{black}{whose state matrix is $\hat{\boldsymbol{A}}$}) and the other representing the dynamics of the rest of the network, as shown in Fig.~\ref{fig_gu_sys_slit}.
For the generator, $u_{qd}^{POC}$ is chosen as an input and $i_{qd}^{tr}$ as an output.
%is chosen to obtain the admittance matrix. 
%
With this input-output selection, the admittance matrix, representing the generator and filter dynamics, is derived by transforming the equations from the state-space form of~\eqref{eq:linear_ode} to the admittance form of~\eqref{eq.imp}, by means of~\eqref{eq.ABCD2imp}.
%
% The system partition is shown in Fig.~(0).
%
%In this figure, a single-generation unit partition from the electrical network is illustrated. 
%Later the unit is split by the system from the \ac{poc} and the conditions for the A matrix are checked. Such conditions are:
%
%The admittance matrix of the generation unit, which contains all of the control and filter dynamics, is produced by taking current as input and voltage as output.
%
Expanding upon~\eqref{eq.imp}, the 2$\times$2 admittance matrix is written as follows
\begin{align} \label{admıttance}
    \begin{bmatrix}
       \Delta i_q^{tr}\\
       \Delta i_d^{tr}
    \end{bmatrix}
    =
        \underbrace{\begin{bmatrix}
       Y_{qq}(s) & Y_{qd}(s)\\
        Y_{dq}(s) & Y_{dd}(s) \\
    \end{bmatrix}}_\text{$\boldsymbol{Y^{sus}(s)}$}
     \begin{bmatrix}
       \Delta u_q^{POC}\\
        \Delta u_d^{POC}
    \end{bmatrix}
\end{align}
where $u_{q,d}^{POC}$ are the $qd$ components of the voltage at the \ac{poc} and $i_{q,d}^{tr}$ are the $qd$ components of the current flowing through the interconnection transformer.
%
%Also, $\boldsymbol{Y^{gu}(s)}$ is the $2 \times 2$  admittance matrix containing the whole control and filter dynamics in the Laplace domain.
%
Each component of the $\boldsymbol{Y^{sus}(s})$ is a \ac{tf}, whose order depends on the filter topology and the complexity of the control.

In order to ensure the validity of the proposed approach, some conditions regarding the eigenvalues of matrix $\hat{\boldsymbol{A}}$ of the system under study need to be ensured. These are the following:
\begin{enumerate}
    \item Matrix $\hat{\boldsymbol{A}}$ has no repeated eigenvalues.
    \item Matrix $\hat{\boldsymbol{A}}$ has no eigenvalues with zero real part.
    \item Matrix $\hat{\boldsymbol{A}}$ has no eigenvalues with positive real part.
\end{enumerate}
Conditions 1 and 2 ensure the dynamics of the system are continuous while condition 3 ensures that the system is stable\textcolor{black}{~\cite{kuznetsov_bifur}}.
One should note that connecting two individually unstable systems can stabilize them and, conversely, separating a stable system to distinct subsystems can render the individual systems unstable~\cite{sun2011impedance}.
%
%separating two connected stable systems might result in individually unstable systems.
%
%Alternatively, when two unstable systems are brought together, they can become stable. 
%
For this reason, it is relevant to check the eigenvalues of the system under study state matrix $\hat{\boldsymbol{A}}$, not the ones of the complete state matrix $\boldsymbol{A}$.

\textcolor{black}{By selecting the desired reference model, detailed in Section~\ref{sec.vsb}, and the frequency range of interest, the methodology is applied as follows.
First, the reference model is subtracted from the system under study in the Laplace domain to derive the $Y^{diff}$ model.
Note that depending on the dynamics under study, the various $Y^{diff}$ models can be obtained.
Such subtraction contains all the dynamic differences between the two models.
Finally, the singular value decomposition (SVD) of the $Y^{diff}$ model is applied over the specified frequency range, where the \ac{dsi} is defined as the maximum singular value.
In the subsequent subsections, the various applications of the methodology will be explained.}
%matrix eigenvalues are checked, not the entire system. 
%
% If the split system is unstable, but a reference is steady. 
% %
% A dynamic comparison between an unstable and stable system can be misleading.
% %
% However, at the system level (will be studied in Sections~\ref{sec.case9 bus} and ~\ref{sec.case118 bus} ), the system is never divided. 
% %
% Basically, the addition of "fictitious current sources" to perform an impedance scan for each bus is done (Section ~\ref{sec.weakspot}).
%
%As a result, condition 3 is significant for generating unit split but not for identification of buses with low VSB.
% It is important to note that condition 3 is necessary to ensure the stability of the generation unit subsystem. 
%check only for generation unit split. For the whole network is not necessary.
%
% This results in the $2 \times 2$  matrices (admittance or impedance), depending on the input-output selection (Step~4 of Fig.~\ref{fig_flow}).
% %
% These matrices contain
% all the control and filter dynamics, as explained in Section~(0).
%
% Later the $2 \times 2$  complex matrix is calculated for each frequency. Note that, the same procedure is done for the reference case hence,
% two $2 \times 2$  complex matrices for each frequency are obtained.
% %
% Taking the Frobenius norm of such matrices at each frequency provides the dynamic similarity across a desired frequency range. 
% %
% Finally, the methodology terminates by taking the average error and normalizing it to achieve the DSI.
%
% \par The system split (Step 4) can be observed in Fig. \ref{fig_gu_sys_slit} for condition checking.
%
\subsection{Component Level Application}
In this section, the application of the \ac{dsi} for the component level is discussed. Once the conditions that are explained above are met, the process to calculate the proposed \ac{dsi} is the following.
First, matrix $\boldsymbol{Y^{sus}(s)}$ is obtained by selecting the corresponding inputs and outputs in the Laplace domain (see~\eqref{admıttance}).

% evaluated at different sampling frequencies by substituting $s$ by $j\boldsymbol{\omega}$ in~\eqref{admıttance}, thus yielding
% %
% \begin{align}
%     \boldsymbol{Y^{gu}}(j \boldsymbol{\omega})
%     =
%     \begin{bmatrix}
%        Y_{qq}(j\boldsymbol{\omega}) & Y_{qd}(j\boldsymbol{\omega})\\
%         Y_{dq}(j\boldsymbol{\omega}) & Y_{dd}(j\boldsymbol{\omega}) \\
%     \end{bmatrix},
% \end{align}
% %
% where $\boldsymbol{\omega}$ is a vector that contains the frequency points of interest.
%
% The length and values of $\boldsymbol{\omega}$ depends on the frequency range of interest for each application. Once the complex matrix is calculated, a matrix called $\boldsymbol{Y^{gu}_{mag}}$, which takes into account the magnitudes, can be written as follows,
% \begin{align}
%     \boldsymbol{Y^{gu}_{mag}}(\boldsymbol{\omega})
%     =
%     \begin{bmatrix}
%        |Y_{qq}(j\boldsymbol{\omega})| & |Y_{qd}(j\boldsymbol{\omega})|\\
%         |Y_{dq}(j\boldsymbol{\omega})| & |Y_{dd}(j\boldsymbol{\omega})| \\
%     \end{bmatrix},
% \end{align}
% where $| \cdot |$ is representing the magnitude of the complex number.
%
%After iterating the $w$ for a defined range of frequencies, the $2 \times 2$  complex matrix is achieved for each frequency.
%
%The same methodology is applied to the reference case which is \ac{vsbi} hence there are two $2 \times 2$  complex matrices for each frequency calculated. 
%
The same process is performed for the reference transfer function $\boldsymbol{Y^{ref}}(s)$, which represents the desired \ac{vsbi} and is defined in~\eqref{eq:vsbi}.
This results in two dynamic systems, one representing the dynamics of the system under study and one representing the desired \ac{vsbi} dynamic characteristics.
%
% \textbf{Remark 1.} \textit{The DSI is the distance between a reference and a generation unit, calculated at each frequency, so, the lower the DSI, the higher the similarity.} \\
%
In order to calculate the dynamic difference between the two systems, the matrices are subtracted and an error matrix $\boldsymbol{Y^{diff}}$ is achieved in the Laplace domain as follows
\begin{equation}
\boldsymbol{Y^{diff}}(s)=\boldsymbol{Y^{ref}}(s)
-
\boldsymbol{Y^{sus}}(s)
\end{equation}
By applying Singular Value Decomposition (SVD) \cite{sigurdskogestad_2005_multivariable} to $\boldsymbol{Y^{diff}}$, the maximum ($\sigma_{max}$) singular values are calculated.
Finally, the \ac{dsi} is defined as
\begin{equation}
DSI(\boldsymbol{\omega})=\sigma_{max}(\boldsymbol{Y^{diff}}(\boldsymbol{\omega}))
\end{equation} 
where $\boldsymbol{\omega}$ is a vector that contains the frequency points of interest.
Based on its definition, \textcolor{black}{the \ac{dsi} represents the maximum amplification of the difference between two systems}.
Smaller \ac{dsi}s indicate more dynamically similar systems, as the \textcolor{black}{difference} is not significantly amplified. On the other hand, larger \ac{dsi}s demonstrate higher \textcolor{black}{difference}, implying that the systems are less dynamically close.
%
% %Also, if the $\sigma_{min}$ is small enough the \ac{dsi} 
% becomes the Frobenius norm of the reference and system under study which is element-wise subtraction of the reference and case study matrices.
% So, \ac{dsi}1 consists of $\sigma_{max}$ calculated per frequency and various signal norms such as 1-norm, 2-norm, and $\infty$-norm can be applied for design purposes.
%
It should be noted that the selection of the frequency range does not affect the accuracy of the \ac{dsi}, provided that the conditions of Section~\ref{sec.dsi_overview} are met and that the linear model accurately represents the full dynamics of the system.
%as long as the conditions mentioned in Section \ref{sec.dsi_overview} are met, and the linear model is considered accurate, the complexity of the models and the frequency range for which \ac{dsi} is evaluated, do not affect its accuracy.
%
%The linear model validation and fulfilling the conditions that are mentioned is critical for the accuracy of the DSI. So, as long as the conditions are met and the linear model is validated the frequency range and the complexity doesn't affect the calculation.\\
%
% \textbf{Remark 2.} \textit{The reference is a user-defined dynamic system.} \\
%
Finally, one should note that the matrix reference $\boldsymbol{Y^{ref}}$ is arbitrary and can be a user-defined dynamic system.
For the purposes of this paper, the reference is a \ac{vsbi}, as defined in~\eqref{eq:vsbi}, but various dynamic models can be selected depending on the application of interest.
\subsection{System Level Application}
\label{sec.weakspot}
In this section, the adaptation of the \ac{dsi} to perform system-wide studies is presented.
%
% The physical meaning of a bus with low \ac{vsb} is the following.
%
By selecting a \ac{vsbi} \textcolor{black}{with either a relatively high or low \ac{scr}*, (where "*" stands for the reference)}, the \ac{dsi} can be calculated for each bus of a power system and for a range of frequency.
Higher \ac{dsi} values indicate lower dynamic similarity between the system under study and the selected reference, as the $\sigma_{max}$ of the error matrix $\boldsymbol{Y^{diff}}$ increases.
When using a high \ac{scr}*, representing a strong grid, a large \ac{dsi} value suggests that the bus is dynamically far from the reference, \textcolor{black}{implying low or very high voltage stiffness} at that bus.
Conversely, when a low \ac{scr}* is used, indicative of a weak grid, \textcolor{black}{large \ac{dsi} values imply high or very low}  voltage stiffness for the bus.
%
% Assuming an arbitrary device to be connected to said bus, the similarity~(or lack thereof) of the dynamics of the rest of the system to a \ac{vsbi} signifies high~(low) \ac{vsb}.
% %
% Alternatively, if the dynamics of the system, as seen by said bus, can be well represented by a \ac{vsbi}~(Thevenin equivalent), then this bus presents high \ac{vsb}.
% %

\begin{figure}[!t]
\centering
\includegraphics[width=1\columnwidth]{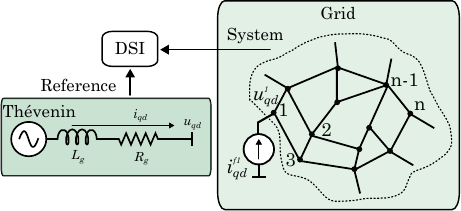}
\vspace{-0.3cm}
\caption{Fictitious current source addition to each bus and comparison of the system with the reference by using DSI}
\label{fig_weak_spot}
\end{figure}

Fig.~\ref{fig_weak_spot} illustrates the methodology to properly define the impedance matrix from the point of view of each bus. In order to achieve the formulation of~\eqref{eq.imp} for each point in the network, a fictitious current source is added to each bus.
This current source injects virtual current in a $qd$ reference frame with the voltages in the $qd$ reference frame being measured at each bus of the network.
 With this method, a $2 \times 2$  impedance matrix consisting of \acp{tf} that contains all the network elements and control dynamics, as seen from each specific bus, \textcolor{black}{is obtained for each bus}.
 Then, by selecting a desired reference (e.g., a \ac{vsbi} with a given \ac{scr}), the \ac{dsi} can be calculated for a range of frequencies.
 %
% Thus, the \ac{dsi} is obtained for each bus with high values for the \ac{dsi} indicating  low \ac{vsb}.
%
%Moreover, $i_{qd}^{f1}$ is the fictitious current in $qd$  injected to bus 1, and $u_{qd}^1$ is the measured voltage in $qd$ at bus 1. Such a thing repeated $n$ times where is the total number of buses. Furthermore, by injecting current and getting voltage, the impedance matrix can be written as follows
Mathematically, the above can be expressed as
\begin{align} \label{impedance}
    \begin{bmatrix}
       \Delta u_q^{b}\\
       \Delta u_d^{b}
    \end{bmatrix}
    =
        \underbrace{\begin{bmatrix}
       Z_{qq}^b(s) & Z_{qd}^b(s)\\
        Z_{dq}^b(s) & Z_{dd}^b(s) \\
    \end{bmatrix}}_\text{$\boldsymbol{Z^{sys,b}(s)}$}
     \begin{bmatrix}
       \Delta i_q^{fb}\\
       \Delta i_d^{fb}
    \end{bmatrix}
\end{align}
where $b \in n$ is the number of buses and $f$ stands for fictitious.
$\boldsymbol{Z^{sys,b}(s)}$ is the $2 \times 2$  impedance matrix in the Laplace domain.
%that contains all the dynamics of the system including network and converter control. 
% Substituting $s=j\boldsymbol{\omega}$ into \eqref{impedance} taking the magnitude for the reference and the system then by subtracting them,
The transfer function of the reference model ($\boldsymbol{Z^{ref}}(s)$) can also be written in a similar way which represents the desired dynamic characteristics. Later, by subtracting  $\boldsymbol{Z^{ref}}(s)$ and $\boldsymbol{Z^{sys,b}}(s)$ in the Laplace domain the $\boldsymbol{Z^{diff}}(s)$ can be written as follows
% the $\boldsymbol{Z^{sys}(s)}$ can  be rewritten as follows
% \begin{align}
%     \boldsymbol{Z^{sys}}=
%     \begin{bmatrix}
%        Z_{qq}(j\boldsymbol{\omega}) & Z_{qd}(j\boldsymbol{\omega})\\
%         Z_{dq}(j\boldsymbol{\omega}) & Z_{dd}(j\boldsymbol{\omega}) \\
%     \end{bmatrix}
% \end{align}
% Again iterating the $\boldsymbol{\omega}$, the $2 \times 2$  complex matrix can be calculated per frequency. Finally taking the Frobenius norm of matrices, \eqref{fro_z} and \eqref{dsı_z} can be written.
% \begin{align} \label{fro_z}
%     ||\boldsymbol{Z^{ref}} - \boldsymbol{Z^{sys}}||_F &= \sqrt{\sum_{x=1}^{c}\sum_{y=1}^{c} |Z^{ref} _{xy} - Z^{sys}_{xy}|^2}
% \end{align}
\begin{align} \label{dsı_z}
   \boldsymbol{Z^{diff,b}}(s) =\boldsymbol{Z^{ref}}(s) - \boldsymbol{Z^{sys,b}}(s)
\end{align}
 Hence the $DSI^b$, that is the \ac{dsi} calculated for bus $b$ can be written as follows,
\begin{align} \label{dsı_z}
   DSI^b(\boldsymbol{\omega})=\sigma_{max}(\boldsymbol{Z^{diff,b}}(\boldsymbol{\omega}))
\end{align}
where $\boldsymbol{Z^{diff,b}}$ is the impedance error matrix calculated for each bus $b$. Finally, the \ac{dsi} values calculated for each bus and frequency point can be written in a matrix form as follows
%vector is normalized with respect to the maximum calculated \ac{dsi} and, can be written as
%
% Finally, the \ac{dsi} vector contains all the computed \ac{dsi}s for each bus can be written as
\[
\boldsymbol{DSI^{b}}(\boldsymbol{\omega})=
\begin{bmatrix}
DSI^1(1) & DSI^2(1) & \cdots & DSI^b(1) \\
DSI^1(2) & DSI^2(2) & \cdots & DSI^b(2)  \\
\vdots & \vdots & \ddots & \vdots \\
DSI^1(\omega)  & DSI^2(\omega) & \cdots & DSI^b(\omega)
\end{bmatrix}_{[\omega \times b]}.
\]
% \begin{align}
%    DSI(\boldsymbol{\omega})=[DSI^1(\boldsymbol{\omega}) DSI^2(\boldsymbol{\omega}) \cdots DSI^n(\boldsymbol{\omega})]_{1\times n}
% \end{align}
%where $\rightarrow$ is the vector indicator.
%
\section{Application Cases of DSI}
\label{sec.case_study}
In this section, different applications of the \ac{dsi} are demonstrated. Section~\ref{sec.control_scheme} describes the control systems that are used in the remainder of the paper for the \ac{gfor} and \ac{gfol} control configurations.
In Section~\ref{sec.case1}, the \ac{dsi} is applied for the \ac{vsb} evaluation of converters with different control topologies and parameter settings in a radial system comprised of a single converter.
%connected with an ideal grid via an $LC$ filter and a transformer.
%
%Noteworthy to show the applicability of the \ac{dsi} for different input/output selections, the admittance matrix is used to demonstrate different behaviors of \ac{gfor} and \ac{gfol} mode with various tunings.
In Sections~\ref{sec.case9 bus} and~\ref{sec.case118 bus}, the proposed methodology is utilized for identifying the buses with low \ac{vsb} at the system level.

For the scope of this paper, the frequency ranges of study are selected as follows:
% Noteworthy, to show the applicability of the \ac{dsi}, the admittance matrix for demonstrating the different behaviors of \ac{gfol} and \ac{gfor} with various tuning and, the impedance matrix for weak spot diagnosis used. 
% \begin{itemize}
%     \item Range 1: 0-20 Hz - Slow interaction converter-driven stability
%     \item Range 2: 20-10 Hz - Low frequency oscillation
%     \item Range 3: 40-100 Hz - Converter-driven stability
%     \item Range 4: 100-1000 Hz - Fast interaction converter-driven stability
% \end{itemize}
\begin{itemize}
    \item Range 1: 0-20 Hz - Slow interaction
    \item Range 2: 20-40 Hz - Low frequency 
    \item Range 3: 40-100 Hz - Converter-driven 
    \item Range 4: 100-1000 Hz - Fast interaction 
\end{itemize}
This range selection was chosen due to the current understanding and classification between dynamic phenomena in power systems and their timescale separation~\cite{freq_range}.
%recent discussions in the literature
%
%However, depending on the study of interest, other frequency ranges can also be considered.
Note that different \ac{dsi} can be computed for different frequency ranges, resulting in different assessments for the \ac{vsb} characteristics of the bus in each frequency range.
%
% In this work, for representing the DSI for a frequency range, the maximum DSI that is inside range is taken. 
%
%For a small case study where the different VSbI behavior of \ac{gfol} and \ac{gfor} with various tuning is shown performed by admittance. Moreover, for the big case study where the weak spot diagnosis is shown applied by impedance.

%Finally, an application of the \ac{dsi} for weak spot diagnosis in terms of voltage will be shown in IEEE 9 bus system.
%
\subsection{Converter Control Topologies: \ac{gfor} and \ac{gfol}}
\label{sec.control_scheme}
This section presents the \ac{gfor} and \ac{gfol} control topologies considered in this work.
%, including \ac{gfor} and \ac{gfol} operation modes.
%
It should be noted that the methodology for the \ac{vsb} assessment presented in this paper is not limited to these control topologies, but can be applied to any control configuration.
The presented topologies are used as representative cases of the two converter control categories.
Both topologies are implemented in a $qd$-reference frame and in a per unit base system.
\begin{figure}[!t]
\includegraphics[width=\columnwidth]{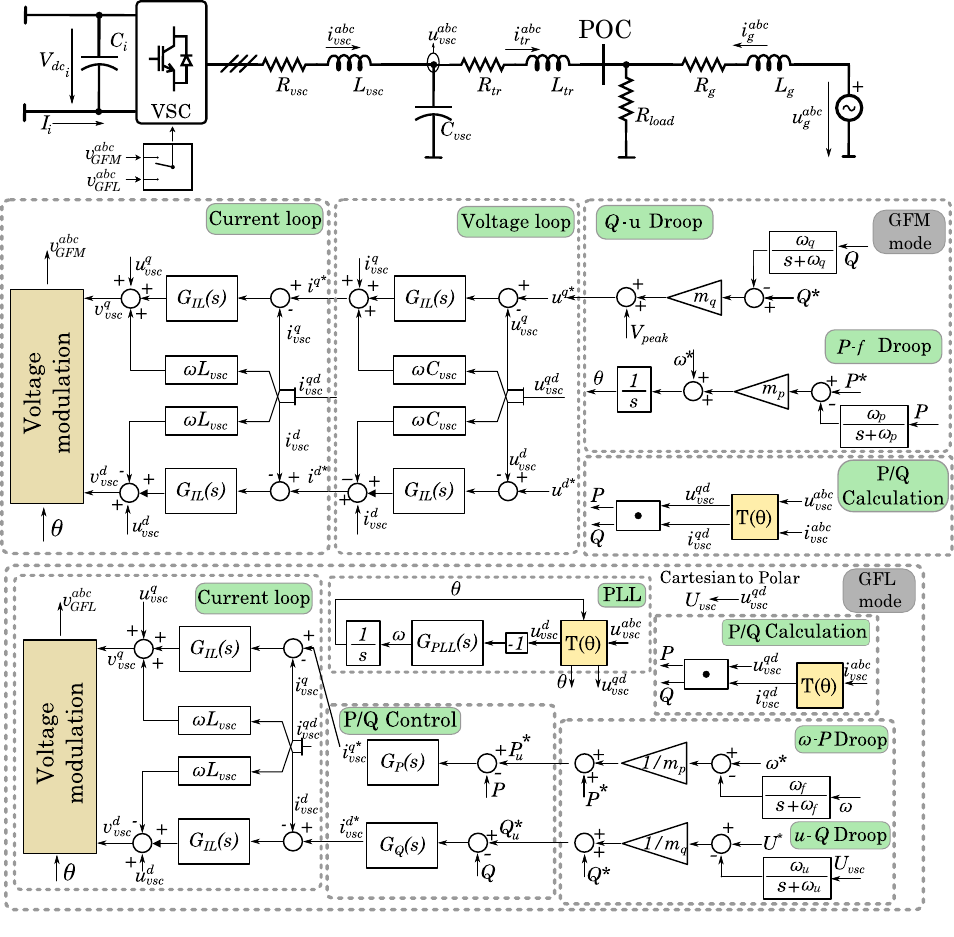}
\caption{\ac{gfol} and \ac{gfor} control topologies connected to a Th\'{e}venin grid via LCL filter topology}
\label{fig:gfol}
\end{figure}

Fig.~\ref{fig:gfol} shows the \ac{gfol} and \ac{gfor} control modes for a converter interfaced with the grid via an $LC$ filter and a connection transformer.
For both configurations, an ideal DC voltage source is assumed on the DC side.
\ac{gfol} converters use a \ac{pll} to synchronize with the network and to provide the angle for internal Park transformations.
\ac{gfol} control is also equipped with $f-P$ and $u-Q$ droops for frequency and voltage support, respectively.
The droop controllers consist of a low-pass filter with cutoff frequencies $\omega_f$ and $\omega_u$. Droop gains of $m_p$ and $m_q$ are used for $f-P$ and $u-Q$ droops, respectively.
These droop controllers provide the active and reactive power references that are used as input signals for the cascaded power and current controllers.
%
%will be used in active and reactive power control (P/Q control).
%
%The P/Q control provides the current references that are then used in the current control to provide the modulation voltage for the converter average model. 
%
Both inner loops use PI controllers that are denoted with $G$ in Fig. \ref{fig:gfol}.
%In this paper, \ac{vsm} is considered as a representative \ac{gfor} controller. 
%
\par For the \ac{gfor} converters, a $P-f$ droop is used to synchronize with the grid and to provide the converter angle for the internal Park transformations.
$m_p$ is the droop coefficient and $\omega_p$ is the cutoff frequency of the first-order filter.
The $Q-u$ droop consists of a low-pass filter with 
 a cutoff frequency $\omega_q$ and droop coefficient $ m_q$ and provides the voltage references to the inner cascaded voltage and current controllers.
%
%The current control is then connected with the voltage control in a cascaded configuration. 
%
%The blocks that are denoted by $G$ are the PI controllers.
%
The interconnection transformer is modeled as a series inductor $L_{tr}$ and resistor $R_{tr}$.
$X_{vsc}$, $R_{vsc}$, and $C_{vsc}$ are the reactance, resistance, and capacitance of the converter filter respectively.
Finally, $R_g$ and $L_g$ are the grid resistance and inductance, respectively.
%
% \begin{figure}[!t]
% \centering
% \includegraphics[width=1\columnwidth]{DSI images v2/fig6_gfol_gfor_dsi_PAPER2.pdf}
% \vspace{-0.3cm}
% \caption{DSI plot of  \ac{gfol} and \ac{gfor} operation modes}
% \label{dsi_gfol_vs_gfor}
% \end{figure}
% %
% \begin{figure}[!t]
% \centering
% \includegraphics[width=0.98\columnwidth]{DSI images v2/bode_gfor_gfol.pdf}\vspace{-0.3cm}
% \caption{Admittance Bode diagram of \ac{gfol}, \ac{gfor} converter and Reference}
% \label{bode_gfol_vs_gfor}
% \end{figure}
% \subsection{Control Performance Evaluation}
% \label{sec.case1}
% Here we showcase a case study for which the \ac{dsi} helps with the tuning of an individual converter. The example can include \ac{gfor}, \ac{gfol} or both.
% %
% Sensitivity analysis of how control parameters affect \ac{dsi}. (presentation can be done with bode etc.)
% %
% Validation through small-signal tools and time domain simulations.
%
\subsection{Component Analysis: Control Performance Effect on the \ac{vsbi}}
\label{sec.case1}
The objective of this subsection is to illustrate that by using the \ac{dsi}, \textcolor{black}{the \ac{vsbi} behaviour can be observed and evaluated for various tuning options of the generating units.}
%
%Note that, the objective is not to reveal which operation mode and which control parameter selection provides better \ac{vsb}, but to showcase how the \ac{dsi} captures the impact of the controllers.
 %Depending on the choice of reference, the interpretation of having more or less \ac{vsb} can change.
 %
 For the study, \ac{vsbi} is selected as a reference with a \ac{scr}* of 15 and X/R ratio of 10.
 First, the \ac{gfol} and \ac{gfor} modes with various control tunings are compared with a \ac{vsbi}  to assess their \ac{vsb}.
 %by using the \ac{dsi}.
 Later, the time domain responses will be shown using simulation to reflect how the \ac{dsi} calculations in the frequency domain translate to the time domain.
\begin{table}[!t]
\vspace{-0.3cm}
  \centering
  \caption{Case study parameters}
    \begin{tabular}{c c c}
    \hline
    \multicolumn{1}{c}{\multirow{2}[2]{*}{Parameters}} & \multicolumn{1}{c}{\multirow{2}[2]{*}{\ac{gfor}}} & \multicolumn{1}{c}{\multirow{2}[2]{*}{\ac{gfol}}} \\
          &       &  \\
    \hline \hline 
    $S_{base}$ & 100 MVA &  100 MVA \\
    
    $V_{base}$ & 230 kV &  230 kV \\
    
    $X_{vsc}$ & 0.15 pu & 0.15 pu \\
    
    $R_{vsc}$ & 0.005 pu & 0.005 pu \\
    
    $C_{vsc}$  & 0.15 pu & 0.04 pu \\
    
    $R_{tr}$   & 0.002 pu & 0.002 pu \\
    
    $X_{tr}$   & 0.1 pu & 0.1 pu \\
   
    $\tau_{PLL}$   & -    & 0.1 s \\
    
    $\tau_{CC}$    & 0.001 s & 0.001 s \\
    
    $\tau_{VC}$    & 0.01 s & - \\

    $\tau_{PQ}$    & - & 0.2 s \\
    
    $m_q$ & 0.02 (pu/pu) & 50 (pu/pu) \\
    
    $m_p$ & 0.05 (pu/pu)  & 20 (pu/pu) \\
    
    $\omega_p$, $\omega_q$ & 50, 10    & -, -\\
    
    $\omega_f$, $\omega_u$ & -, -   & 50, 50\\
 
    \hline
    \end{tabular}%
  \label{tab:casesmall}%
\end{table}
Table~\ref{tab:casesmall} shows the parameters of the case study.
%
%, where $X_{vsc}$, $R_{vsc}$, and $C_{vsc}$ are the reactance, resistance, and capacitance of the converter filter respectively.
%$R_{tr}$ and  $X_{tr}$ are the transformer resistance and reactance.
%
Parameters $\tau_{PLL}$, $\tau_{CC}$, and $\tau_{VC}$ denote the bandwidth of the PLL, current control, and voltage control loops respectively.
\begin{figure}[!t]
\centering
\includegraphics[width=1\columnwidth]{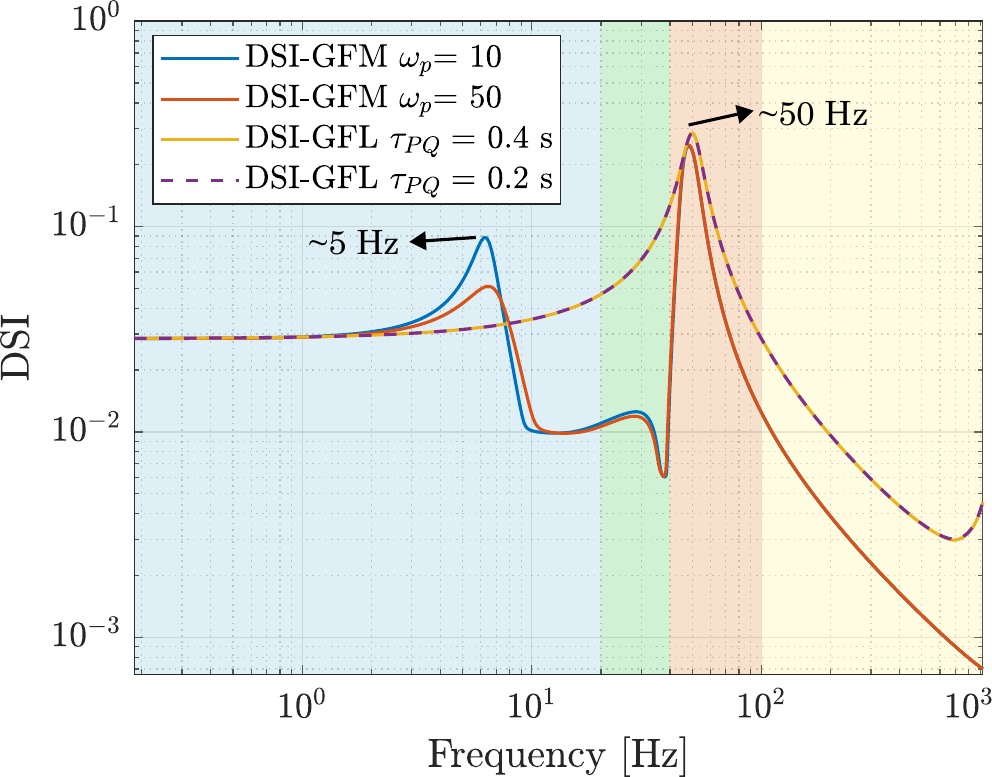}
\caption{Comparison of different tunings of \ac{gfol} and \ac{gfor} with \ac{dsi} plot}
\label{fig_gfolgfor_tuning}
\end{figure}
\begin{figure}[!t] 
\centering
\includegraphics[width=0.95\columnwidth]{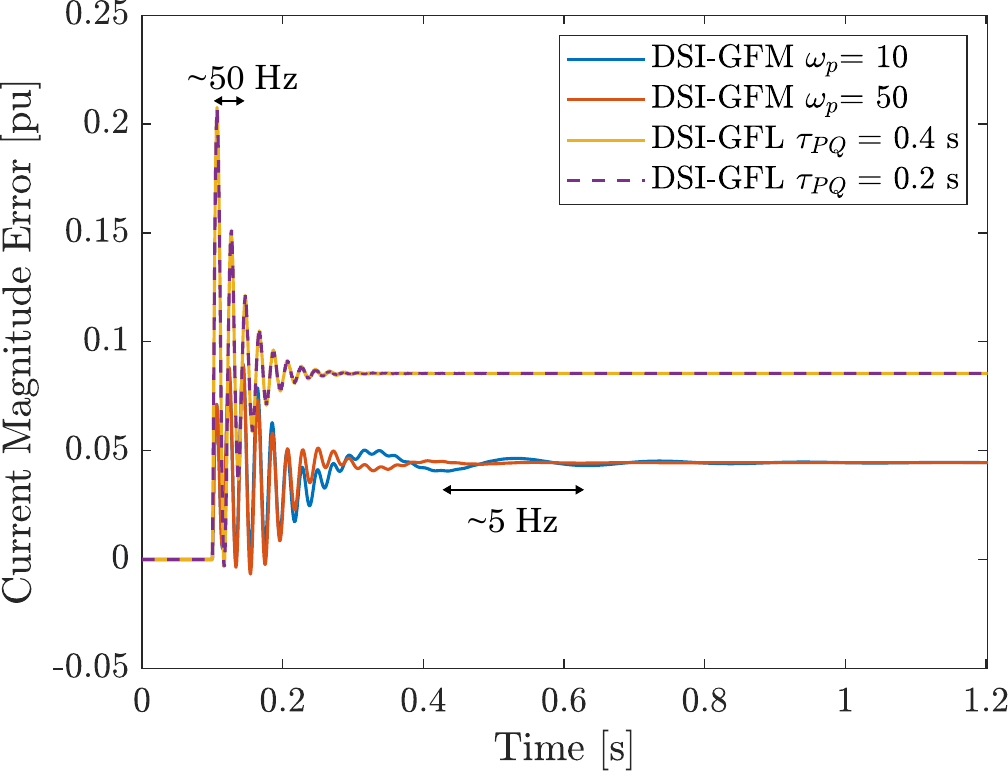}
\caption{Voltage step of 1\% of the peak voltage at the POC. Current magnitudes errors of different tuning of \ac{gfor} and \ac{gfol}. }
\label{td_gfol_gfor}
\end{figure}
% \usepackage{graphicx}
%
For the \ac{gfol} case, the closed loop time response of the outer PQ control loops $\tau_{PQ}$ is changed from 0.2 s to 0.4 s.
For the  \ac{gfor}, the cutoff frequency ($\omega_p$) of the $P-f$ droop is decreased from 50 to 10 while all the rest of the parameters are kept constant.

Fig.~\ref{fig_gfolgfor_tuning} shows the effect of the different parameter selection on the \ac{dsi} plot for both controllers.
Since the reference selected as an \ac{scr}* of 15 and X/R 10 (which corresponds to strong \ac{vsbi}), it can be seen that \ac{gfor} has more similarity to the selected reference than \ac{gfol}.
\textcolor{black}{This is becasue} the \ac{gfor} controller exhibits \textcolor{black}{lower \ac{dsi}, so} smaller error with regards to the strong \ac{vsb} of the reference, compared to the \ac{gfol} one.
%
%However, \ac{gfor} when $\omega_p$ is selected 10 an increase in the \ac{dsi} can be observed around 6 Hz.
%
The selection of the $\omega_p$ parameter has a significant effect around 5 Hz.
%
%Such a thing indicates that the error is getting higher around 6 Hz.
%
This indicates an error increase in the specific frequency area.
Changing the $\tau_{PQ}$ of \ac{gfol} has negligible effect on the dynamics.
Around the operation area of 50~Hz, the \ac{gfol} controller exhibits higher \ac{dsi} and thus, higher error with respect to the reference, compared to the \ac{gfor} controller.
\textcolor{black}{Therefore}, it can be concluded that the \ac{gfor} controller is more similar dynamically than the \ac{gfol} one to a strong voltage source across the majority of the inspected frequency range.
%all the frequency range of study.
%
%However, the parameter selection has an important effect on the error.
%
This is consistent with the current understanding of the two different converter operation modes.
The \ac{dsi} provides the additional benefit of quantifying the error for the different controllers.

Fig.~\ref{td_gfol_gfor} represents the current magnitude errors with respect to the reference after a 1\% POC voltage perturbation.
%when the voltage of the POC increased 1\%.
%
% Note that, the reference is the implementation of the \ac{vsbi} to the network and applying the same disturbance.
 %
It can be observed that \ac{gfol} exhibits a higher error than \ac{gfor}.
%as predicted with the previous analysis
%also shown in Fig. \ref{fig_gfolgfor_tuning}.
%
For \ac{gfor} with $\omega_p$ set to 10, a 5 Hz oscillation is noticeable, and the error is smaller compared to \ac{gfor} with $\omega_p$ at 50.
The steady-state error of \ac{gfol} is higher than \ac{gfor}.
%which is also represented in Fig. \ref{fig_gfolgfor_tuning}.
%
It can be seen that the time domain simulation results are in good agreement with the previous \ac{dsi} analysis.
%
%Finally, it can be said that, with this reference selection and design of \ac{gfol} and \ac{gfor}, \ac{gfor} when $\omega_p$ at 50 has the highest dynamic similarity to the reference along other designs.
 
 % The goal of presenting these time domain findings is not to align them with \ac{dsi} results.
 %
% Rather, the aim is to highlight the dynamic differences in the time domain across the various contexts.
% The GFOR converters have a fast response and changing the $\omega_q$ doesn't affect the response and such a thing could be observed with the DSIs (for all ranges the DSI is close). On the other hand, GFOL converters have a slower and oscillatory behavior. Increasing the $\tau_{PQ}$ from 0.2 s to 0.4 s creates a small difference in the DSIs and such thing can be observed. Such a thing can be observed in the time domain as well when $\tau_{PQ}$ is set to 0.4 s the signal is more damped than 0.2 s. However, in comparison to reference the behaviour of GFOL is different compared to GFOR. Also, the GFOL converters don't have the capability of keeping the voltage at the desired level.
%
% \begin{figure}
% \centering
% \includegraphics[width=1\columnwidth]{9case/td_buses.pdf}
% \vspace{-0.3cm}
% \caption{Voltage of bus 2, 3 and 5 with 1\% of load step in Scenario 1}
% \label{9bus_td}
% \end{figure}
%
%
% \begin{figure}[!t]
% \centering
% \includegraphics[width=1\columnwidth]{9case/td_bus5_v2.pdf}
% \vspace{-0.3cm}
% \caption{Voltage of bus 5 with 1\% resistive disturbance Scenario 2}
% \label{9bus_td_sc2}
% \end{figure}
%
\begin{figure}[!t]
\centering
\includegraphics[width=0.8\columnwidth]{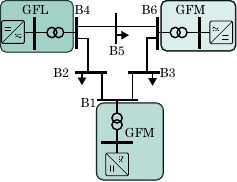}
\caption{Modified IEEE 9 Bus system with the addition of \ac{gfol} and \ac{gfor}}
\label{9bus}
\end{figure}
\begin{figure}[!t]
\centering
\includegraphics[width=1\columnwidth]{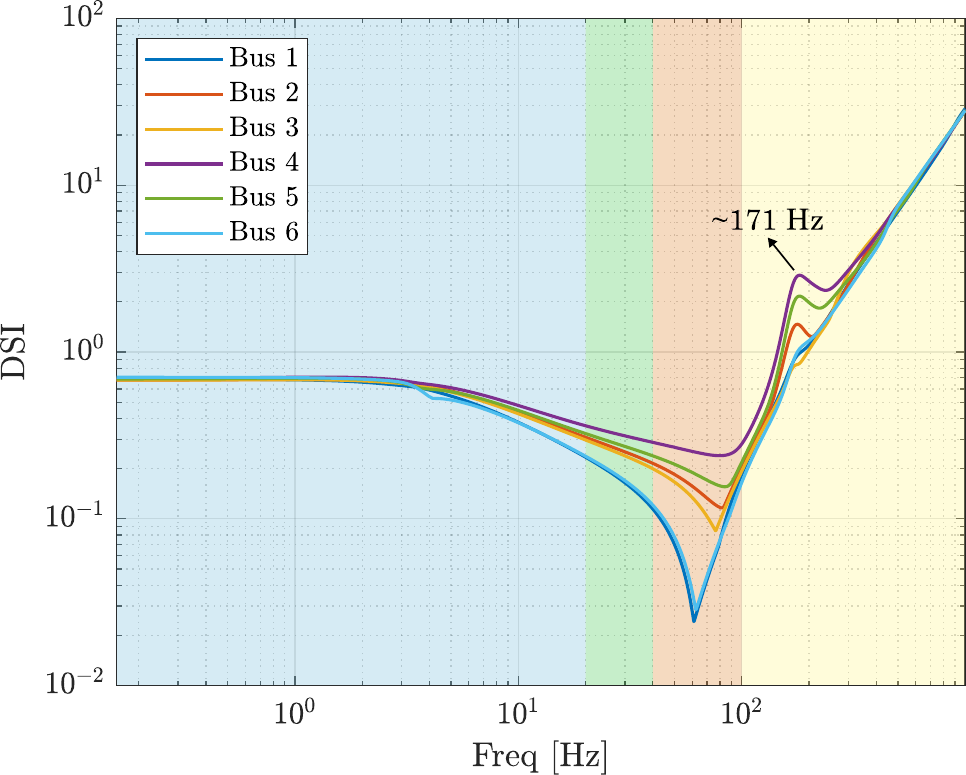}
\caption{\ac{dsi} plot of modified IEEE 9 Bus system \ac{dsi} calculated per bus}
\label{9bus_dsi}
\end{figure}
\begin{figure*}[!t]
%\vspace{-1cm}
\includegraphics[width=\textwidth]{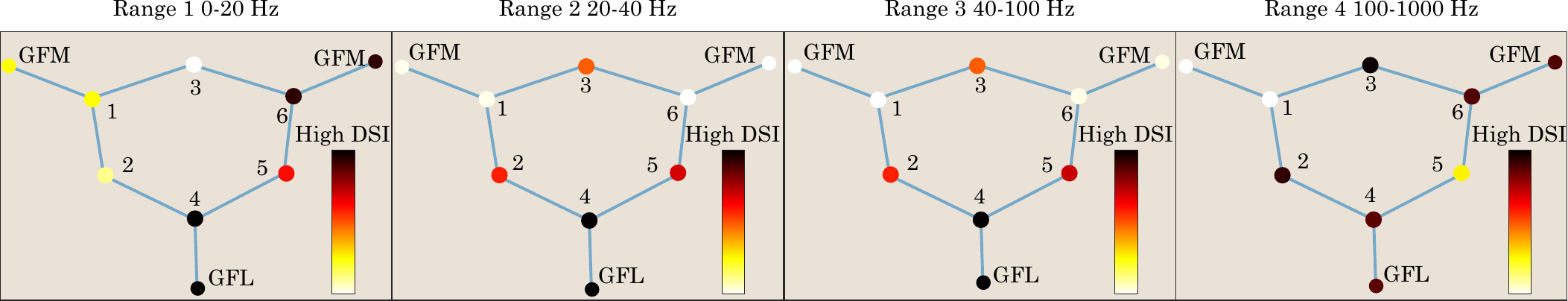}
\vspace{-0.3cm}
\caption{Modified IEEE 9 Bus system with identified buses that have high \ac{dsi} calculated per bus and represented for each range.}
\label{9bus_case}
\end{figure*}

\subsection{Grid Analysis: Application for IEEE 9 Bus System}
\label{sec.case9 bus}
\textcolor{black}{The objective of this subsection is to depict that when DSI is applied at the system level, the buses that exhibit low \ac{vsbi} behaviour can be identified.}
The application of the \ac{dsi} to the system level is demonstrated with a modified version of the benchmark IEEE 9 bus system, depicted in Fig. \ref{9bus}.
%
% There are 3 converters, \ac{gfor} converter is assigned bus 1 (B1) and bus 6 (B6) and to bus 4 (B4) \ac{gfol} converter is selected.
% %
% Moreover, each converters are connected to the network via transformers.
% %
% Since in this paper generation units have $LCL$ filter type, the original 9-bus system is reduced to 6-bus just for representation purposes.
% %
% This modification doesn't affect the original dynamics and steady-state condition of the system.
%
Each bus is denoted with the letter B and, the system has a base apparent power of 100 MVA and a base voltage of 230 kV.
A 310 MVA-rated \ac{gfor} converter is connected to the slack bus B1.
%
%B1 is the Slack bus and where a \ac{gfor} is connected with a nominal power of 310 MVA. 
%
One \ac{gfol} and one \ac{gfor} converters are the other generation units connected to B4 and B6 with a nominal power of 280 and 260 MVA, respectively.
%
%Finally, buses B2, B3 and B5 have PQ loads. 
%
For the dynamic reference, a \ac{vsbi} with a \ac{scr}* of 15 with an $X/R$ ratio of 10 is selected, representing a relatively strong grid.
However, depending on another study of interest another reference can be selected.
% In Table \ref{tab:9bus scenarios} the scenarios are represented.
% \begin{table}[!h]
% \vspace{-0.3cm}
% \centering
% \caption{IEEE 9 Bus system scenarios}
% \label{tab:9bus scenarios}
% \begin{tabular}{c c c c}
% \hline
%            & VSC 1   & VSC 2   & VSC 3   \\ \hline \hline
% Scenario 1 & \ac{gfor} &  \ac{gfol} &  \ac{gfor} \\ \hline
% Scenario 2 &  \ac{gfor} &  \ac{gfol} &  \ac{gfol} \\ \hline
% \end{tabular}
% \end{table}

% First, \ac{gfor} mode is assigned to VSC 1 and VSC 2 while VSC 3 is operated in \ac{gfol} mode with a total \ac{gfol} penetration of 30\%.
% %
% Later, the \ac{gfol} penetration is approximately doubled by assigning VSC 2 as \ac{gfol}. 
% %
% First, $G1$ and $G3$ use GFOR converters, whereas $G2$ uses GFOL converters later, GFOL is replaced by $G3$. For each scenario, each load is disturbed with 1\% and the voltage of that bus is observed.
Fig.~\ref{9bus_dsi} represents the \ac{dsi} calculated for each bus, by using the methodology presented in Section~\ref{sec.weakspot}. In Range 1, the \ac{dsi} values are similar between the different system buses.
In Ranges 2 and 3, significant variation can be observed, as buses 1 and 6 have a lower \ac{dsi} than the rest of the buses and bus 4 has the highest.

This is consistent with the system topology as buses 1 and 6 have \ac{gfor} converters connected to them, which demonstrate higher similarity to the selected reference in these frequency ranges.
Accordingly, in such ranges \ac{gfol} indicate higher dynamic error, which is consistent with the analysis and results of Section~\ref{sec.control_scheme}.
Also, a peak in the \ac{dsi} can be observed in bus 4 and 5 around 171 Hz indicating lower dynamic similarity at such frequency. 
Finally, at higher frequencies, the system \ac{dsi} is increasing, hence reflecting an overall trend of reduced dynamic similarity to the reference.

Fig.~\ref{9bus_case} displays the modified IEEE 9 bus system as a graph for each frequency range, with a color code identifying the different values of \ac{dsi} that are used in the graph. 
Note that for the graph presentation, the \ac{dsi} values in each range are normalized based on the maximum and minimum \ac{dsi} within that range. 
This normalization is applied solely for observation purposes, allowing independent evaluation of each frequency range.
%
%Subsequently, the full frequency spectrum of the \ac{dsi} is presented to provide a comprehensive view of the system's dynamic behavior.
%
% The interpretation of the color code is the following.
% %
% The higher the calculated \ac{dsi}, the darker the bus color leading to a dynamically distant behavior from the selected reference and vice versa.
%
% Since, in this work, the selected reference is a strong grid (\ac{scr} 15 and X/R ratio 10), being dynamically distant is interpreted as having low \ac{vsb}.
%

In Range 1, bus 4 exhibits the highest \ac{dsi} value, followed by buses 6 and 5, indicating that their voltage dynamics show the least similarity to the selected reference.
Conversely, buses 1, 2, and 3 display lower values, suggesting a higher degree of dynamic similarity to the reference.
%when compared to Buses 1, 4, and 5.
%
For Range 2, bus 4 showcases the highest \ac{dsi} value, indicating the lowest similarity, followed by buses 2 and 5.
%
%In contrast, Buses 3, 2, and 5 exhibit lighter colors, reflecting a greater alignment with the reference dynamics. 
%
In Range 3, bus 1 presents the lowest value, suggesting the highest similarity to the reference, followed by buses 6 and 3. 
%
%Buses 4, 5, and 2, however, show darker shades, indicating less similarity.
%
Finally, for Range 4, bus 1 has the lowest value, signifying the closest match to the reference, while the remaining buses have higher values.
%
%Note that, in Ranges 2,3 and 4 Bus 1 and Bus 6 which have \ac{gfor} have brighter colors than Bus 4 which has a \ac{gfol} converter.
%
%Such a thing aligns with the results that are achieved in Section \ref{sec.control_scheme}.
%
The above results provide insight about the voltage stiffness in the different buses of the network, based on the topology of the system and the dynamic operation of the interconnected devices, which can be leveraged for the isolation of problematic buses and guide corrective actions.
% %
% \begin{figure}[!t] %% 9 bus td results whole thevenin replacement
% \centering
% \includegraphics[width=0.97\columnwidth]{9case/9bus_td.pdf}
% \caption{Voltage magnitude errors of generation buses (1, 4, and 6) with 1\% load disturbance of bus 5 including magnifying graph.}
% \label{9bus_td_ind}
% \end{figure}
% %
%
\begin{figure}[!t] %% 9 bus td results whole thevenin replacement
\centering
\includegraphics[width=0.97\columnwidth]{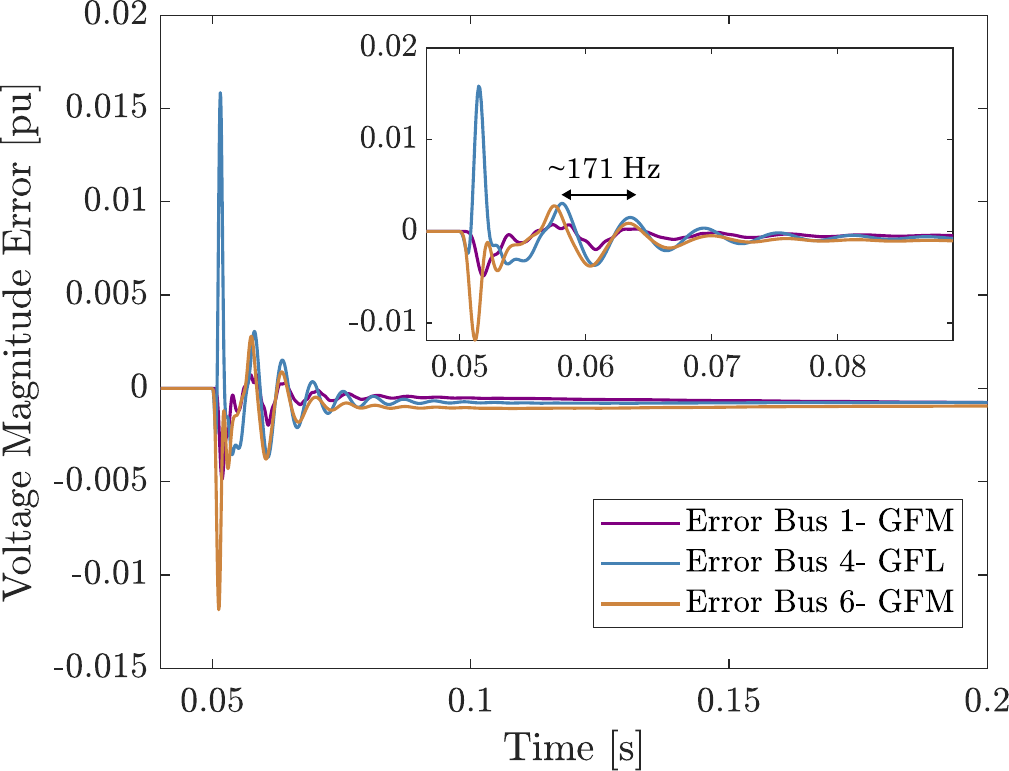}
\caption{Voltage magnitude errors of generation buses (1, 4, and 6) with 1\% load disturbance of bus 5 including magnifying graph.}
\label{9bus_td_ind}
\end{figure}
Fig.~\ref{9bus_td_ind} represents the voltage magnitude errors of buses 1, 4 and 6 with 1\% of load change in load 5.
For the error calculation, a reference system was used for which the converter-based generators were \textcolor{black}{individually replaced by Th\'{e}venin circuit} with the same impedance characteristics that were used for the \ac{dsi} reference calculation.
In Fig. \ref{9bus_td_ind} when the load at bus 5 is incremented by 1\%, \ac{gfol} exhibits the largest error.
Notably, the voltage error of bus 1, which has \ac{gfor}, is more damped than bus 4.
Oscillation around 171 Hz can be observed especially in the error of bus 4 which is reflected in Fig.~\ref{9bus_dsi} with an increment of \ac{dsi}.
Finally, the time domain simulation results are matching with the \ac{dsi} analysis shown in Fig.~\ref{9bus_dsi}.

\begin{figure}[!t]
\centering
\includegraphics[width=1\columnwidth]{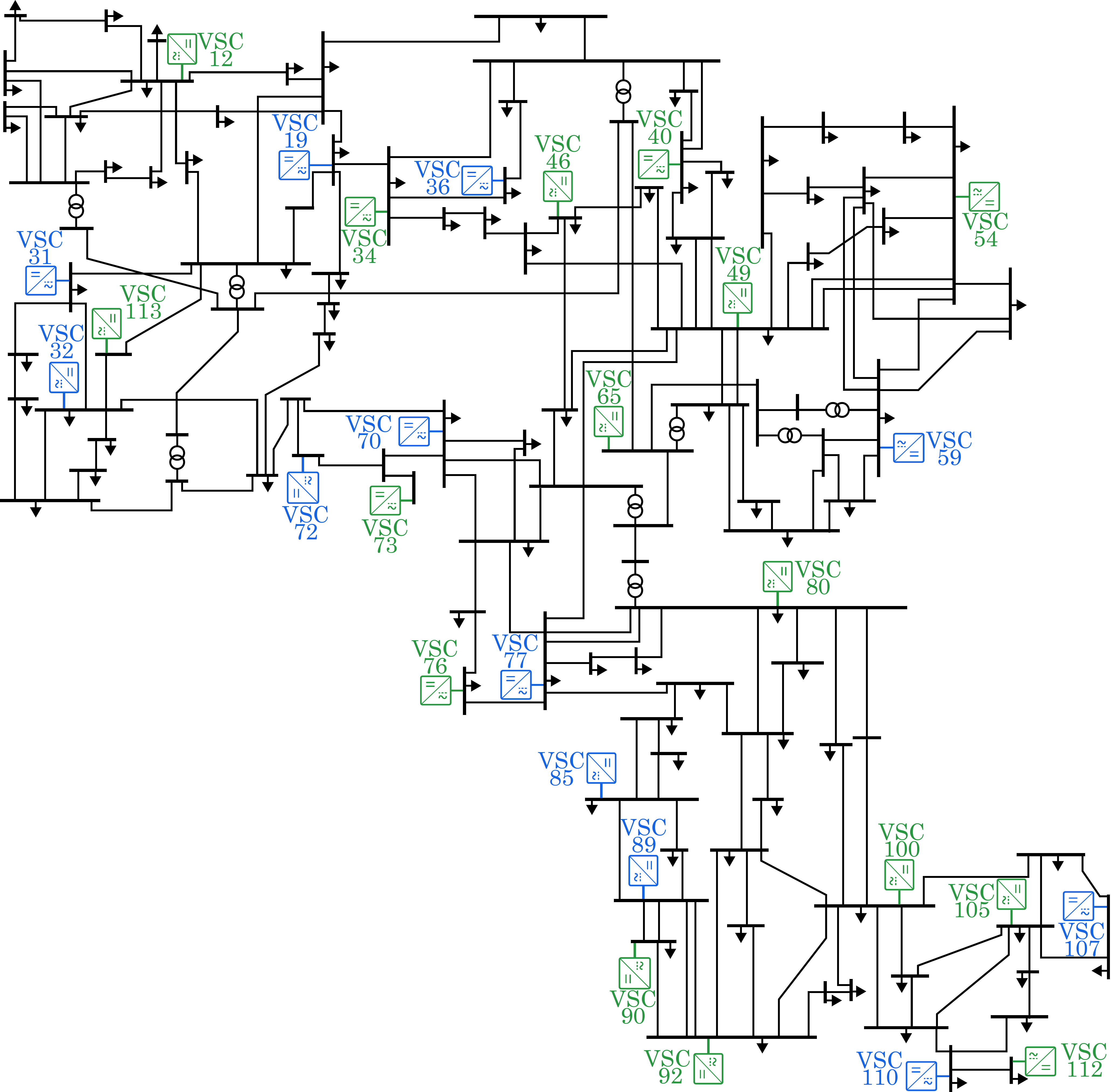}
\vspace{-0.3cm}
\caption{ Modified IEEE 118 Bus system including 12 \ac{gfol} (blue) and 16 \ac{gfor} (green)}
\label{118_scheme}
\end{figure}
\begin{figure*}[!t]
\centering
\includegraphics[width=\textwidth]{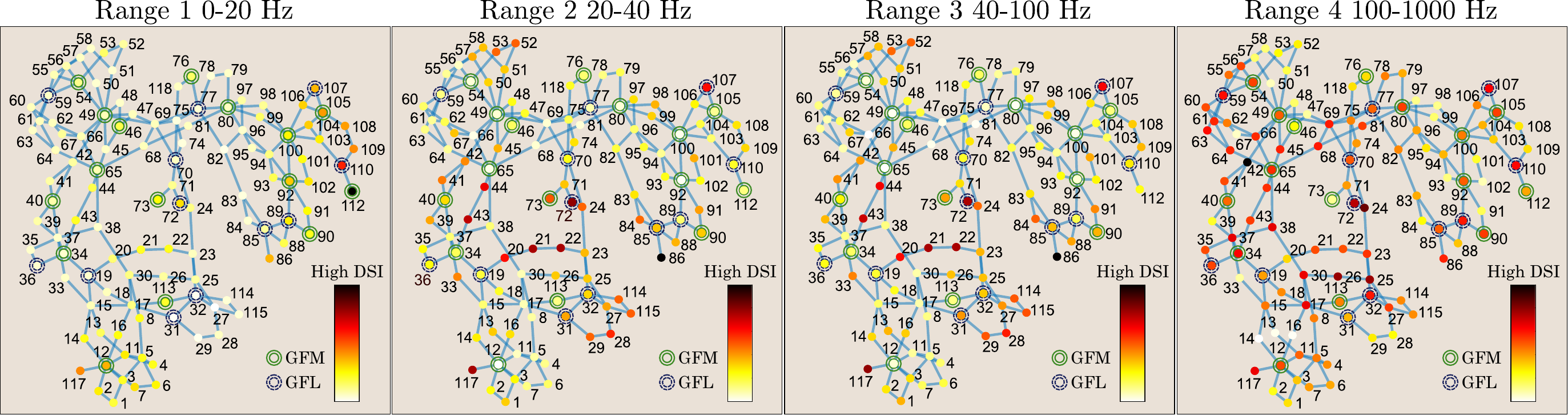 }
\vspace{-0.5cm}
\caption{ Modified IEEE 118 Bus system including 12 
\ac{gfol} and 16 \ac{gfor} converters with identified buses that have high \ac{dsi} calculated per bus and represented for each range.}
\label{118_weakpoints}
\end{figure*}
% %
%%%----%% REMOVED PLOTS 
% \begin{figure}[!ht]
% \centering
% \includegraphics[width=1\columnwidth]{118case/dsi_plot_118.pdf}
% \vspace{-0.3cm}
% \caption{\ac{dsi} plot of modified 118 bus system of buses 17, 86 and 117}
% \label{118_dsi}
% \end{figure}
% %
% \begin{figure}[!ht]
% \centering
% \includegraphics[width=0.97\columnwidth]{118case/td_118.pdf}
% \vspace{-0.3cm}
% \caption{The error of the voltage magnitudes of Bus 17,86 and 117 with respect to the reference including magnifying graph}
% \label{118_td}
% \end{figure}
% %
%
\subsection{Grid Analysis: Application for IEEE 118 Bus System}
\label{sec.case118 bus}
\textcolor{black}{The objective of the subsection is to show the scalability of the proposed method and the calculation efficiency of the system-wide \ac{dsi}, the same analysis was applied to a modified version of the well-known IEEE 118 bus system.}
%
%In this case, the application of the \ac{dsi} to identify the buses with low VSB will be applied to the modified IEEE 118 bus system.
%
The system under study is shown in Fig.~\ref{118_scheme}, with 12 generation units operating in \ac{gfol} mode (in blue), and 16 operating in \ac{gfor} mode (in green).
The total \ac{gfol} penetration is 28.6~\% with respect to the total nominal power.
%
%Moreover, in order to apply the methodology, a \ac{vsbi} with \ac{scr} of 15 and $X/R$ 10 is selected as a reference.
%
For the \ac{dsi} reference, the same impedance parameters were used as in the previous cases while similar normalization was used as in the 9 bus case.
%that was used for this case is similar with the one of the 9 bus case.
%
% Noteworthy, for the graph presentation, the \ac{dsi} values in each range are normalized based on the maximum and minimum \ac{dsi} within that range. 
% %
% This normalization is applied solely for observation, allowing independent evaluation of each frequency range.
% %
% Later, the full frequency spectrum of the \ac{dsi} is presented to provide a comprehensive view of the system's dynamic behavior.
%
The number of states was increased to 1159 states from 73, compared to the 9 bus case.

Fig.~\ref{118_weakpoints} presents the system as a graph with the \ac{dsi} values for each bus and for the different frequency regions of the study.
%the identified buses are shown for different frequency ranges.
%
%The interpretation of the color code is the same as in Section \ref{sec.case9 bus}.
%
For Range 1, buses 112, 110, and 117 display higher \ac{dsi} values indicating a dynamically distant behavior compared to the selected reference for the rest of the buses.
However, buses such as 31, 32, 27, and 29 have low \ac{dsi} values which show the most dynamic similarity compared to the selected reference.
For Range 2 and 3, bus 86 has the highest \ac{dsi} followed by 72 and 117, implying the biggest error for the selected reference and, the rest of the buses have low \ac{dsi} values.
Finally, for Range 4, bus 42 has the highest \ac{dsi}, however, buses such as 13, 14 and 16 have low \ac{dsi} values showing more similarity to the selected reference than the rest of the buses.
So, considering the above results, the modified IEEE 118 Bus system with such topology and selected reference, buses 42, 72, 86, and 117 has the lowest similarity to a \ac{vsbi} with a \textcolor{black}{SCR*} of 15 and $X/R$ of 10, indicating that such buses are more susceptible to voltage variations from the rest of the buses. 
%
%Note that, with various control and location of \ac{ibr}, and reference selection various outcomes can be obtained.
%
The results from the above analysis can be used for the identification of problematic parts of the network and for resiliency enhancing, system-level planning.
%in order to enhance the system durability.

%
%Since \ac{dsi} is a flexible tool and can be applied to various grid topologies with different control implementations and reference selection, it is crucial to represent the computational time of the method.
%
\begin{table}[!t]
\vspace{-0.3cm}
  \centering
  \caption{Computation time of the IEEE 9 and 118 Bus systems}
    \begin{tabular}{c c c}
    \hline
     &  \begin{tabular}[c]{@{}c@{}} 9 Bus \\ (73 states) \end{tabular}& \begin{tabular}[c]{@{}c@{}} 118 Bus \\ (1159 states) \end{tabular} \\
    \hline \hline 
    \begin{tabular}[c]{@{}c@{}}State-space calculation\end{tabular} & 7 s &  9 s \\
    Condition checking & 0.2 s &  39.3 s \\
    \ac{dsi} calculation & 0.5 s & 3616 s \\
    Total & 7.7 s & 3664.3 s \\
    \hline
    \end{tabular}%
  \label{tab:time_}%
\end{table}

Table \ref{tab:time_} shows the necessary computation time for applying the proposed method for the 9 and 118 bus cases, divided between each necessary step (see Fig. \ref{fig_flow}).
The state-space calculation is done automatically in Matlab taking into account the topology of each system, hence no notable difference is observed for the two systems.
%for building 9 or 118 bus systems.
%
The necessary time to perform the condition checking, detailed in Section \ref{sec.dsi}, of 118 bus system is 196 times higher than the 9 bus case.
%however, it is still 39.3 s which can be considered as relatively fast.
%
% Among the other steps of the methodology, the \ac{dsi} calculation is the most time-consuming.
%
For the 9 bus \ac{dsi} calculation, 0.5 seconds were required to calculate the DSI while for the 118 bus system, 3616 seconds ($\sim$60.21 minutes) were required.
%
%Since there is no simplification of the network or control topology, getting fully accurate results for 118 bus systems lasts 1110.3 seconds which is approximately 18.5 minutes.
%
One should note that computation time can vary significantly depending on the control design complexity, network size, and frequency resolution which, for this study, was set to a step of 0.15 Hz for a range of 0-1000 Hz.
The calculations were performed on a 13th Gen Intel(R) Core(TM) i7-1355U 1.70 GHz processor with 40 GB RAM.

\section{Conclusion}\label{sec.conc}
This paper proposes a dynamic similarity index that quantifies the \textcolor{black}{difference dynamic responses of two systems.}
%
%
%provides an analytical derivation of the voltage source behind an impedance in the frequency domain.
%
This index is calculated for a range of frequencies by including all the desired dynamics.
\textcolor{black}{In order to apply the \ac{dsi}, a reference model is selected for the comparison against a desired dynamic response. 
In this work, a \ac{vsbi} is selected as a reference and two applications of the \ac{dsi} are performed by using such reference.
It is shown that, applying DSI to a single component evaluates its dynamic performance against a desired reference, reflecting, in this particular case, the VSB of the converter control. At the system level, DSI diagnoses buses with low VSB, providing critical insights into grid strength, currently defined as SCR}.
% The applicability of the \ac{dsi} is flexible since any dynamic system can be selected as a reference.
%
% Then, the \ac{dsi} was applied to assess the voltage source behavior of different operation modes (\ac{gfor} and \ac{gfol}) of power converters with different control parameter selection.
% %
% By using \ac{dsi}, the buses of transmission networks that exhibit low \ac{vsb} can be determined.
%
% The theoretical developments and the presented analysis are supported with different case studies based on modified versions of the IEEE 9 and 118 bus systems with 100\% converter penetration.
%
% It is shown that the \ac{dsi} can be used at the component level, for evaluating the \ac{vsb} of the converter control and, at the system level, offers insights into grid strength, currently defined as \ac{scr}.}
%as well as of different buses in large networks.%and also can be applied to a system level. 
%
%Finally, the \ac{dsi} can also be applied for control design and system-level planning which can be useful for consultancy companies and TSOs.
%
%Future work can include, a quantified \ac{dsi} which can give more insight about the interpretability of \ac{dsi} and expand it for large signal studies.
%
Future work will focus on defining alternative references \textcolor{black}{to present \ac{vsb} or other dynamic responses, beyond the conventional \ac{vsbi}} for the \ac{dsi} calculation, as well as on optimizing the control setup of a converter to minimize its dynamic error from a desired reference.
%for various dynamics and points for the network. 
%
%Later, the proper design applications will be done for converter-based resources based on \ac{dsi}.
%
% Knowing what level of \ac{vsbi} generation units can provide depending on the operation mode and tuning can be observed with DSI which can be useful for system-level or control design.
% %
% Finally, identifying buses that have low \ac{vsb} by using DSI can be beneficial for \ac{tso}s in order to provide network requirements.
%
\appendices
% \section{Validation}
% \begin{figure}[!h]
% \centering
% \includegraphics[width=1\columnwidth]{DSI images/validation.pdf}
% \caption{Validation of GFOL and GFOR converters 1\% voltage change in the Thevenin grid}
% \label{valid_gfol_gfor}
% \end{figure}
\section{Analytical derivation of damping ratio}
\label{sec.appen1}
%
%In this paper, the $X/R$ ratio has been kept constant at 10 for the reference case since the scope is in the transmission level. However, 
The analytical relationship between impedance parameters $R$ and $L$ and design specifications $X/R$ and $\xi$ (damping ratio) is derived in this Appendix.
%can be derived.
%
%By knowing the second-order behavior \eqref{eq:vsbi} $F(s)$ 

Starting from the second order \ac{tf} of equation~\eqref{eq:vsbi}
% %
% \begin{align}
%     F(s)={s^2+\frac{2sR}{L}+\frac{R^2+\omega^2L^2}{L^2}}
% \end{align}
% %
and by setting $k=R/L$, considering the second order expression $as^2+bs+c$, the following relations are derived
\begin{align}
    a=1, \; b=2k, \; c=k^2+\omega^2.
\end{align}
The discriminant $D$ can be computed as follows
\begin{align}
    D=b^2-4ac, \; \sqrt{D}=-j2\omega.
\end{align}
The poles can be calculated as follows
\begin{align}
    x_{1,2}
    &=\frac{-b\pm\sqrt{D}}{2a}\\
    &=-k+j\omega.
\end{align}
The damping factor $\xi=\frac{k}{\sqrt{k^2+\omega^2}}$ can be rewritten as follows
\begin{align} \label{eq:dam}
    \xi\sqrt{(k^2+\omega^2)}=k.
\end{align}
Taking the square of \eqref{eq:dam} and solving for $k$ we derive
\begin{align} \label{eq:dam1}
    k=\frac{\xi \omega}{\sqrt{1-\xi^2}}.
\end{align}
Substituting back $k=\frac{R}{L}$ into~\eqref{eq:dam1} 
\begin{align}
    \frac{R}{L}=\frac{\xi \omega}{\sqrt{1-\xi^2}}.
\end{align}
%
% In order to achieve the $X/R$ ratio
% $X=\omega L$ so $L=\frac{X}{\omega}$ so we obtain the analytical expression between $X/R$ and the damping ratio $\xi$:
% %
% \begin{align}
%     \frac{X}{R}=\frac{\sqrt{1-\xi^2}}{\xi}.
% \end{align}
Replacing  $L=\frac{X}{\omega}$ and isolating $\xi$ we can write
\begin{align}
    \xi=\frac{R}{\sqrt{X^2+R^2}}.
\end{align}
\section*{Acknowledgment}
This work was supported by FEDER / Ministerio de Ciencia e Innovación
- Agencia Estatal de Investigación, under the project REFORMING (PID2021-127788OA-I00). The work of O. Gomis was supported by the ICREA Academia program. The work of E. Prieto and M. Cheah was supported by Serra Húnter Program. The work of J. Amorós has been partially supported by the project
PID2023-146936NB-I00 financed by the Spanish State Agency MCIN/AEI,
FEDER, UE.
% \IEEEtriggeratref{8}
% \IEEEtriggercmd{\enlargethispage{-5in}}

\bibliographystyle{IEEEtran}
\bibliography{refs.bib}

\begin{thebibliography}{10}
\providecommand{\url}[1]{#1}
\csname url@samestyle\endcsname
\providecommand{\newblock}{\relax}
\providecommand{\bibinfo}[2]{#2}
\providecommand{\BIBentrySTDinterwordspacing}{\spaceskip=0pt\relax}
\providecommand{\BIBentryALTinterwordstretchfactor}{4}
\providecommand{\BIBentryALTinterwordspacing}{\spaceskip=\fontdimen2\font plus
\BIBentryALTinterwordstretchfactor\fontdimen3\font minus \fontdimen4\font\relax}
\providecommand{\BIBforeignlanguage}[2]{{%
\expandafter\ifx\csname l@#1\endcsname\relax
\typeout{** WARNING: IEEEtran.bst: No hyphenation pattern has been}%
\typeout{** loaded for the language `#1'. Using the pattern for}%
\typeout{** the default language instead.}%
\else
\language=\csname l@#1\endcsname
\fi
#2}}
\providecommand{\BIBdecl}{\relax}
\BIBdecl

\bibitem{kundur}
P.~Kundur, \emph{Power System Stability and Control}.\hskip 1em plus 0.5em minus 0.4em\relax McGraw-Hill Companies, 01 1994.

\bibitem{sallam_2015_power}
A.~A. Sallam and O.~P. Malik, \emph{Power system stability : modelling, analysis and control}.\hskip 1em plus 0.5em minus 0.4em\relax Institution Of Engineering And Technology, 2015.

\bibitem{2017_high}
ENTSO-E, ``High penetration of power electronic interfaced power sources (hpopeips),'' 2017.

\bibitem{2021_global}
IEA, ``Global energy review,'' 2021.

\bibitem{moutevelis2023taxonomy}
D.~Moutevelis, J.~Rold{\'a}n-P{\'e}rez, M.~Prodanovic, and F.~Milano, ``Taxonomy of power converter control schemes based on the complex frequency concept,'' \emph{IEEE Trans. on Pow. Sys.}, 2023.

\bibitem{PAOLONE}
M.~Paolone \emph{et~al.}, ``Fundamentals of power systems modelling in the presence of converter-interfaced generation,'' \emph{Electric Power Systems Research}, vol. 189, p. 106811, 2020.

\bibitem{2019_high}
ENTSOE, ``High penetration of power electronic interfaced power sources and the potential contribution of grid forming converters,'' p.~32, 2019.

\bibitem{pll_weak}
J.~Matevosyan \emph{et~al.}, ``A future with inverter-based resources: Finding strength from traditional weakness,'' \emph{IEEE Power and Energy Magazine}, vol.~19, no.~6, pp. 18--28, 2021.

\bibitem{wang2020grid}
X.~Wang, M.~G. Taul, H.~Wu, Y.~Liao, F.~Blaabjerg, and L.~Harnefors, ``Grid-synchronization stability of converter-based resources—an overview,'' \emph{IEEE Open Journal of Industry Applications}, vol.~1, pp. 115--134, 2020.

\bibitem{rosso2021grid}
R.~Rosso, X.~Wang, M.~Liserre, X.~Lu, and S.~Engelken, ``Grid-forming converters: Control approaches, grid-synchronization, and future trends—a review,'' \emph{IEEE Open Journal of Industry Applications}, vol.~2, pp. 93--109, 2021.

\bibitem{esig}
\BIBentryALTinterwordspacing
``Grid-forming technology in energy systems integration es energy systems integration group,'' 2022, eSIG report. [Online]. Available: \url{https://www.esig.energy/reports-briefs.}
\BIBentrySTDinterwordspacing

\bibitem{gfor}
D.~Rathnayake \emph{et~al.}, ``Grid forming inverter modeling, control, and applications,'' \emph{IEEE Access}, vol.~PP, pp. 1--1, 08 2021.

\bibitem{Zhao2022}
F.~Zhao, X.~Wang, and T.~Zhu, ``Power dynamic decoupling control of grid-forming converter in stiff grid,'' \emph{IEEE Transactions on Power Electronics}, vol.~37, pp. 9073--9088, 8 2022.

\bibitem{Zhao2023}
------, ``Low-frequency passivity-based analysis and damping of power-synchronization controlled grid-forming inverter,'' \emph{IEEE Journal of Emerging and Selected Topics in Power Electronics}, vol.~11, pp. 1542--1554, 4 2023.

\bibitem{Xu2023}
Z.~Xu, N.~Zhang, Z.~Zhang, and Y.~Huang, ``The definition of power grid strength and its calculation methods for power systems with high proportion nonsynchronous-machine sources,'' \emph{Energies}, vol.~16, 2023.

\bibitem{Shah2022}
S.~Shah, P.~Koralewicz, V.~Gevorgian, and R.~Wallen, ``Sequence impedance measurement of utility-scale wind turbines and inverters - reference frame, frequency coupling, and mimo/siso forms,'' \emph{IEEE Transactions on Energy Conversion}, vol.~37, pp. 75--86, 3 2022.

\bibitem{freq_gfol_gfor}
\BIBentryALTinterwordspacing
J.~M. Julia~Matevosyan, ``Grid-forming technology in energy systems integration es energy systems integration group,'' 2022. [Online]. Available: \url{https://www.esig.energy/reports-briefs.}
\BIBentrySTDinterwordspacing

\bibitem{VATTAKKUNI20236042}
K.~{Vatta Kkuni}, S.~Mohan, G.~Yang, and W.~Xu, ``Comparative assessment of typical control realizations of grid forming converters based on their voltage source behaviour,'' \emph{Energy Reports}, vol.~9, pp. 6042--6062, 2023.

\bibitem{vsb_linbin}
H.~Xin, C.~Liu, X.~Chen, Y.~Wang, E.~Prieto-Araujo, and L.~Huang, ``How many grid-forming converters do we need? a perspective from small signal stability and power grid strength,'' \emph{IEEE Transactions on Power Systems}, pp. 1--13, 2024.

\bibitem{modal}
I.~J. Perez-arriaga, G.~C. Verghese, and F.~C. Schweppe, ``Selective modal analysis with applications to electric power systems, part i: Heuristic introduction,'' \emph{IEEE Transactions on Power Apparatus and Systems}, vol. PAS-101, pp. 3117--3125, 1982.

\bibitem{machowski_2011_power}
J.~Machowski, J.~W. Bialek, and D.~J. Bumby, \emph{Power System Dynamics}.\hskip 1em plus 0.5em minus 0.4em\relax John Wiley \& Sons, 2011.

\bibitem{canizares1998calculating}
C.~A. Canizares, ``Calculating optimal system parameters to maximize the distance to saddle-node bifurcations,'' \emph{IEEE Transactions on Circuits and Systems I: Fundamental Theory and Applications}, vol.~45, no.~3, pp. 225--237, 1998.

\bibitem{chen2021improving}
Y.~Chen, R.~Preece, and M.~Barnes, ``Improving system loadability with the consideration of multiple bifurcations,'' \emph{International Journal of Electrical Power \& Energy Systems}, vol. 132, p. 107182, 2021.

\bibitem{ogata2010modern}
K.~Ogata, \emph{Modern control engineering fifth edition}, 2010.

\bibitem{sun2011impedance}
J.~Sun, ``Impedance-based stability criterion for grid-connected inverters,'' \emph{IEEE transactions on power electronics}, vol.~26, no.~11, pp. 3075--3078, 2011.

\bibitem{moutevelis2022bifurcation}
D.~Moutevelis, J.~Rold{\'a}n-P{\'e}rez, M.~Prodanovic, and S.~Sanchez-Acevedo, ``Bifurcation analysis of active electrical distribution networks considering load tap changers and power converter capacity limits,'' \emph{IEEE Transactions on Power Electronics}, vol.~37, no.~6, pp. 7230--7246, 2022.

\bibitem{collados}
C.~Collados-Rodriguez, M.~Cheah-Mane, E.~Prieto-Araujo, and O.~Gomis-Bellmunt, ``Stability analysis of systems with high vsc penetration: Where is the limit?'' \emph{IEEE Transactions on Power Delivery}, vol.~35, no.~4, pp. 2021--2031, 2020.

\bibitem{kuznetsov_bifur}
Y.~Kuznetsov, \emph{Elements of Applied Bifurcation Theory}.\hskip 1em plus 0.5em minus 0.4em\relax Springer Science \& Business Media, 2008.

\bibitem{sigurdskogestad_2005_multivariable}
S.~Skogestad and I.~Postlethwaite, \emph{Multivariable Feedback Control}.\hskip 1em plus 0.5em minus 0.4em\relax John Wiley \& Sons, 11 2005.

\bibitem{freq_range}
N.~Hatziargyriou \emph{et~al.}, ``Definition and classification of power system stability – revisited \& extended,'' \emph{IEEE Transactions on Power Systems}, vol.~36, no.~4, pp. 3271--3281, 2021.

\end{thebibliography}
\end{document}